\documentclass[floatfix,aps,prd,showkeys,twocolumn,nofootinbib]{revtex4-2}

\usepackage{amsmath}
\usepackage{amssymb}
\usepackage{xcolor}
\usepackage{graphicx}
\usepackage{dcolumn}
\usepackage[loop]{animate}
\usepackage{comment}
\usepackage[normalem]{ulem}
\usepackage[bottom]{footmisc}
\usepackage{float}
\usepackage{hyperref}
\usepackage{parskip}

\newcommand{\p}{\partial}
\newcommand{\pnabla}{{}^\perp\nabla}

\newcommand{\const}{\text{const}}
\newcommand{\Ai}{\text{Ai}}

\newcommand{\diag}{\text{diag}}

\begin{document}

\title{Instability windows of relativistic r-modes}

\date{\today}

\author{Kirill Y. Kraav}
\author{Mikhail E. Gusakov}
\author{Elena M. Kantor}
\affiliation{Ioffe Institute, Polytekhnicheskaya 26, St.-Petersburg 194021, Russia}


\begin{abstract}

The detectability of the gravitational-wave signal from $r$-modes depends on the interplay between the amplification of the mode by the CFS instability and its damping due to dissipative mechanisms present in the stellar matter. The instability window of $r$-modes describes the region of stellar parameters (angular velocity, $\Omega$, and redshifted stellar temperature, $T^\infty$), for which the mode is unstable. In this study, we reexamine this problem in nonbarotropic neutron stars, taking into account the previously overlooked nonanalytic behavior (in $\Omega$) of relativistic $r$-modes and enhanced energy dissipation resulting from diffusion in superconducting stellar matter. We demonstrate that at slow rotation rates, relativistic $r$-modes exhibit weaker amplification by the CFS instability compared to Newtonian ones. However, their dissipation through viscosity and diffusion is significantly more efficient. In rapidly rotating neutron stars within the framework of general relativity, the amplification of $r$-modes by the CFS mechanism and their damping due to shear viscosity become comparable to those predicted by Newtonian theory. In contrast, the relativistic damping of the mode by diffusion and bulk viscosity remains significantly stronger than in the nonrelativistic case. Consequently, account for diffusion and general relativity leads to a substantial modification of the $r$-mode instability window compared to the Newtonian prediction. This finding is important for the interpretation of observations of rotating neutron stars, as well as for overall understanding of $r$-mode physics.

\end{abstract}

\maketitle


\section{Introduction}

A perturbed neutron star (NS) can oscillate in many different ways, each distinguished by the geometry of the corresponding fluid element displacements and the restoring force that influences its dynamics the most. Predominantly toroidal oscillations of rotating stars, governed primarily by the Coriolis force, are known as $r$-modes in the literature. Compared to other oscillations, $r$-modes are the most easily \cite{andersson1998, fm1998} driven unstable by the Chandrasekhar-Friedman-Schutz (CFS) mechanism. According to this mechanism \cite{chandra1970, fs1978_1, fs1978_2, friedman1978}, gravitational-wave emission from the perturbed star triggers the transfer of the stellar rotation energy into the oscillation energy, leading to the growth of the oscillation amplitude. Therefore, the strong CFS instability of the $r$-modes makes them promising targets for gravitational radiation searches. If detected, $r$-mode gravitational signal would provide valuable information on the physics of the superdense stellar matter, irreproducible in terrestrial laboratories. Unfortunately, current electromagnetic and gravitational-wave observations only allow to set the upper limits on the $r$-mode amplitude, but the next generation gravitational-wave detectors might be able to capture the first signals from $r$-modes (see, e.g.,  \cite{zhuetal2019, cppo2022, hhcc2019, roci2021, abbotetal2021, boztepeetal2020, bhattacharyaetal2017, carideetal2019} and references therein).

The fact that no reliable $r$-mode detection has occurred so far can be explained not only by the insufficient sensitivity of the gravitational-wave detectors, but also by the influence of various dissipative mechanisms, preventing the mode amplitude from growing. Indeed, an $r$-mode can develop in a star with a given angular velocity, $\Omega$, and redshifted temperature, $T^\infty$, only if the CFS mechanism (whose efficiency depends on $\Omega$) can overcome the mode damping caused by dissipative processes (the latter depend on $\Omega$ and $T^\infty$). The region in the $(\Omega, T^\infty)$ plane, where this condition is satisfied, defines what is known as the $r$-mode instability window \cite{low1998, levin1999, bildstenetal2000, ak2001, cgk2017}.

One of the main problems in the physics of $r$-modes is identification of the most efficient dissipative mechanisms, responsible for the $r$-mode stabilization. According to numerical calculations (see, e.g., \cite{owenetal1998, levin1999, hl2000, haskell14}), a neutron star with excited $r$-mode cannot spend much time deep inside the instability window due to its rapid spin-down by gravitational-wave emission. In other words, the probability to find a star there is negligibly small. Therefore, dissipation should be strong enough to form the instability window, that would not contain observable sources. The failure of the frequently considered ``minimal'' model \cite{low1998}, where the shear viscosity determines the energy dissipation at low and moderate temperatures, while at high temperatures the mode is stabilized by the bulk viscosity, implies the existence of other energy loss channels. A number of extensions of the minimal model, such as dissipation in the Ekman layer \cite{bu2000, lu2001, ga2006}, mutual friction \cite{hap2009}, resonant $r$-mode stabilization by superfluid modes \cite{kgd2020, kgd2021} and enhanced bulk viscosity in hyperonic matter \cite{nb2006, ofengeimetal2019} have been considered in the literature as possible candidates (see reviews \cite{glampedakis2018, haskell14} for details). Here we would like to study another possibility.

In our study, we improve the theory of the $r$-mode instability windows by modifying it in three ways. The first two modifications, accounting for the stellar matter nonbarotropicity and relativity, are tightly related to the so-called ``continuous spectrum problem'': while the traditional perturbation theory in the slow-rotation approximation predicts the existence of the {\it continuous} part in the oscillation spectrum, more accurate treatment beyond this approximation leads to {\it discrete} oscillation frequencies (see the detailed review in \cite{kgk2022_1}). Recently, we have managed to show \cite{kgk2022_1, kgk2022_2}, that the reason for the slow-rotation approximation breakdown is that, in nonbarotropic slowly-rotating neutron stars, relativistic generalization of the Newtonian $r$-modes is described by nonanalytic functions of $\Omega$, which, in turn, results in the relations between the $r$-mode eigenfunctions (their ``ordering''), significantly different from those of the Newtonian theory. As we shall see, the peculiarities of the relativistic $r$-mode eigenfunctions significantly influence the shape of the instability window.

The third modification of the theory we would like to make is to determine the efficiency of particle diffusion in damping the relativistic $r$-modes. Previously we have shown \cite{kgk2021} that diffusive dissipation is significantly enhanced in superconducting stellar matter, and becomes the dominant dissipative mechanism for sound waves, $p$-modes, and $g$-modes. Although in the very same study we found that the effect of diffusion on the $r$-mode instability window is weak in the Newtonian theory, in General Relativity (GR), as we shall see, the situation is completely different and diffusion becomes the leading dissipative mechanism at not too high stellar temperatures.

This paper is structured as follows. Section II provides an overview of key assumptions about the properties of neutron star matter, its equilibrium, modeling of stellar oscillations, and the consideration of the CFS instability and dissipative effects. In that section we also present general expressions for the calculation of the oscillation energy change rates and timescales, associated with shear viscosity, bulk viscosity, diffusion and CFS instability. Section III starts with the brief review of the nonanalytic relativistic $r$-mode properties, followed by the derivation of the $r$-mode energy change rates, associated with evolutionary mechanisms under consideration. We compare the derived expressions with their Newtonian counterparts and consider the limit of the extremely slow rotation, where the difference is especially pronounced. We finish section III by presenting the results of our numerical calculation of the $r$-mode evolutionary timescales and instability windows. Finally, section IV contains a discussion of the results obtained and some concluding remarks.

Throughout the text almost all the equations are written in the dimensionless form, unless stated otherwise (whether we use dimensionless form or not is always clear from the context). This means that by all the quantities in equations we imply their dimensionless counterparts. These counterparts are obtained by measuring mass in units of the stellar mass $M$, distance in units of the stellar radius $R$, and time in units $\Omega_{\rm K}^{-1}$ , where $\Omega_{\rm K}\equiv\sqrt{GM/R^3}$ is a quantity of the order of Keplerian frequency and $G$ is the gravitational constant. Although in the chosen units $G=1$, we still retain $G$ in equations. With $G$ retained, the dimensionless and non-dimensionless forms of {\it any} equation {\it exactly} concide with each other.


\section{Theoretical framework}


\subsection{Stellar matter}

As outlined in the introduction, our aim is to clarify the effects of the peculiar behavior of relativistic $r$-modes and energy dissipation due to particle diffusion on the $r$-mode instability window. Both of these effects are inherent primarily to the neutron star core. Diffusion is important only when enhanced by the strong proton superconductivity in the core \cite{kgk2021}, while the relativistic $r$-mode peculiarities arise due to nonbarotropicity of the core matter (the neutron star crust can be considered as effectively barotropic \cite{kgk2022_1}). We would also like to note that the effects, associated with the presence of the stellar crust, are unlikely to qualitatively change our results (see Sec.\ \ref{IV Conclusion}). For these reasons in what follows, unless stated otherwise, we ignore the crust and use the toy model of a neutron star, consisting only of a liquid nonbarotropic core. In our study, we also neglect the stellar magnetic field, since it is unlikely that it significantly affects the $r$-mode dynamics \cite{chirenti2013, kinney2003}. Finally, dissipation and gravitational radiation effects only weakly affect stellar oscillations, and their influence typically manifests itself on the timescales much larger than the oscillation period. This allows us to model stellar oscillations within the nondissipative hydrodynamics, and account for these effects perturbatively, as discussed in Sec.\ \ref{Sec II D}.

We think of a neutron star as of a liquid mixture of different particle species, labelled throughout the text by Latin indices ($i, j, k,\dots$). We characterize the stellar matter by the total pressure $p$, energy density $\varepsilon$, enthalpy density $w=p+\varepsilon$, temperature $T$, and by the electric charges $e_k$, chemical potentials $\mu_k$ and number densities $n_k$ of different particle species $k$ in the mixture. For simplicity, we restrict ourselves to the case, when all the particle species are normal (i.e., not superfluid or superconducting) except for a one charged completely superconducting component ``$\rm s$" (in reality, we imply protons by this component). Then, in the absence of dissipation, the macroscopic flow of normal particle species $j_k^\mu=n_k \mathfrak{u}^\mu$ is described by the collective ``normal" four-velocity $\mathfrak{u}^\mu$, while that of the superconducting component $j^\mu_{\rm s}=n_{\rm s}\mathfrak{u}_{\rm s}^\mu$ is described by the ``superconducting" four-velocity $\mathfrak{u}_{\rm s}^\mu$. All the mentioned thermodynamic quantities are, by definition, measured in the reference frame, comoving with the ``normal" component of the fluid, which implies that for any particle species $\mathfrak{u}_\mu j_k^\mu=-n_k$ and, therefore, $\mathfrak{u}_\mu \mathfrak{u}_{\rm s}^\mu=-1$ (the summation over repeated Greek indices is implied).
 
In superconducting matter the thermodynamic quantities, generally, depend on the relative velocity $\mathfrak{u}_{\rm s}^\mu-\mathfrak{u}^\mu$. When considering the global stellar dynamics it is, however, a very good approximation to assume that electric charges and electric currents almost cancel each other out, so that the condition $e_k j_k^\mu=0$ (the summation over repeated Latin indices is implied) is met with tremendous accuracy, ensured by the adjusting self-consistent electric field. In our case this condition reduces to the equality $\mathfrak{u}_{\rm s}^\mu=\mathfrak{u}^\mu$, and the dependence of thermodynamic quantities on the relative velocity can be ignored. In degenerate stellar matter one can also ignore their temperature dependence, so that any thermodynamic quantity $f$ can be considered as a function  only of the set of number densities $\{n_k\}$. Given the equation of state (EOS), $\varepsilon=\varepsilon(\{n_k\})$, provided by the microscopic theory, one can use thermodynamic relations 
\begin{gather}
d\varepsilon=\mu_k dn_k, \quad dp=n_k d\mu_k, \quad w=\mu_k n_k
\end{gather}
to find $p(\{n_k\})$, $w(\{n_k\})$, and $\mu_m(\{n_k\})$. We will be interested in the case of nonbarotropic EOS, when these dependencies cannot be parametrized by a single quantity (say, baryon number density). Note, however, that in equilibrium such parametrization becomes possible due to additional conditions, imposed by the chemical equilibrium.

Due to equality $\mathfrak{u}_{\rm s}^\mu=\mathfrak{u}^\mu$, the nondissipative stress-energy tensor $T^{\mu\nu}$ of the described superconducting neutron star effectively coincides with that of nonsuperconducting case
\begin{gather}
T^{\mu\nu}=w \mathfrak{u}^\mu \mathfrak{u}^\nu+p\mathfrak{g}^{\mu\nu},
\end{gather}
where $\mathfrak{g}_{\mu\nu}$ is the metric tensor. As a result, the equations, governing the stellar dynamics and structure, are essentially the same, as in nonsuperconducting case. As for the superconductivity, it becomes important in consideration of the viscous and diffusive dissipation, since it affects viscous and diffusion coefficients, as well as the self-consistent electric field  (to be discussed below).


\subsection{Stellar equilibrium}

According to Hartle \cite{hartle1967}, the spacetime of a neutron star, rotating slowly with angular velocity $\Omega$, is described in the $x^\mu=(ct,r,\theta,\varphi)$ coordinates by the interval of the following form:
\begin{multline}
\label{geometry}
ds^2={\rm g}_{\mu\nu}dx^\mu dx^\nu=-e^{2\nu(r)}c^2 dt^2+e^{2\lambda(r)}dr^2+\\
+r^2\{d\theta^2+\sin^2\theta[d\varphi-\Omega\omega(r)dt]^2\}+O(\Omega^2).
\end{multline}
This geometry possesses two independent Killing vectors: vector $t^\mu=\delta^\mu_t$ is associated with the conserved energy of the matter and vector $\varphi^\mu=\delta^\mu_\varphi$ is associated with the conserved angular momentum. Their linear combination $k^\mu=t^\mu+(\Omega/c)\varphi^\mu$ is also a Killing vector, that can be associated with the conserved energy, as measured in the corotating reference frame. In terms of these vectors the four-velocity, corresponding to the equilibrium geometry \eqref{geometry}, can be written as (further we use the subscript ``$0$'' to denote the equilibrium value $f_0$ of any quantity $f$)
\begin{gather}
u^\mu\equiv\mathfrak{u}_0^\mu=\Lambda[t^\mu+(\Omega/c)\varphi^\mu]=\Lambda k^\mu, \\
\label{Lambda}
\Lambda=(-{\rm g}_{\mu\nu}k^\mu k^\nu)^{-1/2}=e^{-\nu(r)}+O(\Omega^2).
\end{gather}
The condition of thermal equilibrium requires the redshifted temperature, $T^\infty$, to be constant (see, e.g., \cite{mirallesetal1993}):
\begin{gather}
\label{equilibriumT}
T^\infty\equiv T/\Lambda=\const.
\end{gather}
The chemical potentials in the unperturbed star, in turn, should satisfy the relations (e.g., \cite{rjfk06})
\begin{gather}
\label{equilibriumMU}
\p_\rho\biggl(\frac{\mu_{k0}}{\Lambda}\biggr)-\frac{e_k E_{0\rho}}{\Lambda}=0, \quad E^\rho\equiv \mathfrak{u}_\lambda F^{\rho\lambda},
\end{gather}
where $\p_\rho\equiv\p/\p x^\rho$, $F^{\mu\nu}$ is the electromagnetic tensor and $E^\rho$ is the electric four-vector. For normal particle species this condition follows from the absence of entropy generation due to diffusion in equilibrium, while for the superconducting component it follows from the ``superconducting'' hydrodynamic equation (to be discussed below). Finally, note that from these conditions one immediately obtains
\begin{gather}
\label{equilibriumMU2}
e_k\mu_{m0}^\infty-e_m\mu_{k0}^\infty\equiv (e_k\mu_{m0}-e_m\mu_{k0})/\Lambda=\const.
\end{gather}

In what follows we, for simplicity, ignore the oblateness of the neutron star due to rotation, i.e., we ignore $O(\Omega^2)$-terms in the metric tensor \eqref{geometry} and redshift \eqref{Lambda}, and also assume that the equilibrium value $f_0$ of any thermodynamic quantity $f$ depends only on the coordinate $r$. For further convenience in transitioning to the Newtonian limit, we also avoid explicitly expressing the redshift $\Lambda$ in the equilibrium conditions \eqref{equilibriumT}--\eqref{equilibriumMU2} as $e^{-\nu(r)}$ (see Sec.\ \ref{III B} for details).


\subsection{Stellar hydrodynamics}

The most general way to describe perturbations of a neutron star over the equilibrium state is to consider {\it exact} deviation $\hat{\delta}f\equiv f-f_0$ of any physical quantity $f$ in a perturbed star from its equilibrium value, $f_0$. If we, for example, consider the product $(fg)$ of two functions, its full perturbation will be equal to
$\hat{\delta}(fg)=\hat{\delta} f g_0+f_0\hat{\delta}g+\hat{\delta}f \hat{\delta} g$. In case of a weak stellar perturbation, one usually retains only the linear terms in the perturbation amplitude and looks for the approximate solution to the oscillation equations in the form $f\approx f_0+\delta f$, where $\delta f$ is often referred to as the Eulerian perturbation of a quantity $f$. In what follows by the Eulerian perturbation, we denote a quantity linear in the amplitude, which satisfies the Leibnitz rule: $\delta(fg)=\delta f g_0 + f_0\delta g$. The smaller the perturbation amplitude, the less is the error induced by using the Eulerian perturbations instead of the exact perturbations. It is thus natural to use Eulerian perturbations when considering weak perturbations, unless one is interested in quantities, quadratic in the perturbation amplitude, such as oscillation energy in the frame rotating with the star (see Appendix \ref{AppA}).

In our study, we treat small deviations of a neutron star from the equilibrium state within the Cowling approximation (i.e., we ignore metric perturbations in an oscillating star), which greatly simplifies the problem and, at the same time, provides a reasonably accurate estimate of the properties of the global stellar oscillation modes \cite{lsr1990, yk1997, jc2017} (see also the discussion of the Cowling approximation in application to nonanalytic relativistic $r$-modes in \cite{kgk2022_1, kgk2022_2}). Within this approximation, the nondissipative dynamics of small stellar perturbations is described by the following closed system of equations:
\begin{gather}
\label{hydrodynamics}
\left\{
\begin{gathered}
\delta[\mathfrak{u}^\rho\nabla_\rho \mathfrak{u}^\mu+(1/w)\perp^{\mu\rho}\nabla_\rho p]=0 \\
\delta[\nabla_\mu j_k^\mu]=0 \\
\delta\mu_m=(\p\mu_m/\p n_k)_0\delta n_k \\
\delta p=(\p p/\p n_k)_0\delta n_k=n_{k0}\delta\mu_k \\
\delta w=\delta\varepsilon+\delta p=\mu_{m0} \delta n_m+\delta p
\end{gathered}
\right.,
\end{gather}
where $\nabla_\rho$ is the covariant derivative, associated with the geometry \eqref{geometry}, and $\perp^{\mu\nu}\equiv \mathfrak{g}^{\mu\nu}+\mathfrak{u}^\mu \mathfrak{u}^\nu$ is the orthogonal projection tensor. The first equation of the system \eqref{hydrodynamics} is the linearized relativistic Euler equation, $\perp^\mu_\rho \nabla_\lambda T^{\rho\lambda}=0$, the second is the set of continuity equations for different particle species, and the remaining equations follow from the EOS and thermodynamic relations in degenerate matter.

Superconductivity brings to the problem an additional hydrodynamic equation for the velocity $\mathfrak{u}_{\rm s}^\mu$, that can be used to find the self-consistent electric field $E^\mu$. In the absence of flux tubes, the ``superconducting" velocity is defined by the equation \cite{kg11,gkcg13}:
\begin{gather}
\mathfrak{u}_{{\rm s}\mu}=(1/\mu_{\rm s})[\nabla_\mu\phi-e_{\rm s}A_\mu],
\end{gather}
where $A^\mu$ is the electromagnetic four-potential, and $\phi$ is a scalar proportional to the phase of the Cooper-pair condensate wave function. Note that this definition of the superconducting velocity slightly differs from that adopted, e.g., in Ref.\ \cite{gusakov2016}.

The phase of the Cooper-pair condensate wave function should naturally satisfy the potentiality condition, meaning that $\nabla_\mu\nabla_\nu \phi=\nabla_\nu\nabla_\mu \phi$, or, equivalently,
\begin{gather}
e_s F_{\mu\nu}=\nabla_\nu (\mu_{\rm s}\mathfrak{u}_{{\rm s}\mu})-\nabla_\mu(\mu_{\rm s} \mathfrak{u}_{{\rm s}\nu}),
\end{gather}
where we have used the definition of the electromagnetic tensor, $F_{\mu\nu}\equiv \nabla_\mu A_\nu-\nabla_\nu A_\mu$. Using that in our case $\mathfrak{u}_{{\rm s}}^\mu=\mathfrak{u}^\mu$ and contracting this equation with $\mathfrak{u}^\nu$, one finds:
\begin{gather}
\label{Efield0}
E_\mu=(1/e_{\rm s})[\mathfrak{u}^\nu\nabla_\nu(\mu_{\rm s} \mathfrak{u}_\mu)-\mathfrak{u}^\nu\nabla_\mu(\mu_{\rm s}\mathfrak{u}_\nu)].
\end{gather}
Using the explicit form of the four-velocity, $u^\mu=\Lambda k^\mu$, and the Killing equation, $\nabla_{\mu}k_\nu+\nabla_\nu k_\mu=0$, one can show that in equilibrium this equation, as promised, reduces to the condition \eqref{equilibriumMU}, written for the superconducting component. Finally, with the use of the Euler equation for $\mathfrak{u}^\mu$ and thermodynamic relations, we rewrite \eqref{Efield0} as
\begin{gather}
\label{Efield}
E_\mu=\frac{\Lambda}{e_{\rm s}}\biggl[\delta_{{\rm s}k}-\frac{\mu_{\rm s} n_k}{w}\biggr]\pnabla_\mu\biggl(\frac{\mu_k}{\Lambda}\biggr), \ \pnabla_\mu\equiv\perp_\mu^\rho\nabla_\rho.
\end{gather}

Instead of dealing with the four-velocity perturbation, $\delta\mathfrak{u}^\mu$, we find it more convenient to operate with the Lagrangian displacement vector $\xi^\mu$, showing the variation of the fluid element world lines, induced by a perturbation. In the Cowling approximation, the velocity perturbation and Lagrangian displacement are related as (see, e.g., \cite{friedman1978})
\begin{gather}
\delta\mathfrak{u}^\mu=-\perp^\mu_\rho\mathcal{L}_\xi u^\rho=\perp^\mu_\rho(u^\lambda\nabla_\lambda\xi^\rho-\xi^\lambda\nabla_\lambda u^\rho),
\end{gather}
where $\mathcal{L}_\xi$ is the Lie derivative along the vector field $\xi^\mu$. The invariance under the gauge transformation $\xi^\rho\to \xi^\rho+f u^\rho$ with arbitrary function $f$ can be used to impose an additional condition $u_\mu\xi^\mu=0$. It is also convenient to replace the angular components of the displacement with the functions $\mathrm{Q}$ and $\mathrm{T}$, defined as
\begin{gather}
\begin{gathered}
\xi^\theta=\frac{1}{r}\biggl[\frac{\p \mathrm{Q}}{\p\theta}+\frac{1}{\sin\theta}\frac{\p \mathrm{T}}{\p\varphi}\biggr], \\
\xi^\varphi=\frac{1}{r \sin\theta}\biggl[\frac{1}{\sin\theta}\frac{\p \mathrm{Q}}{\p\varphi}-\frac{\p \mathrm{T}}{\p\theta}\biggr].
\end{gathered}
\end{gather}
Once $\mathrm{Q}$ and $\mathrm{T}$ are expanded in spherical harmonics, such representation becomes equivalent to that used by Regge \& Wheeler \cite{rw1957} and Thorne \& Campolattaro \cite{tc1967}. Function $\mathrm{T}$ will further be referred to as the {\it toroidal function} (not to be confused with the temperature, $T$).

In terms of the Lagrangian displacement and Lagrangian perturbations, $\Delta\equiv\delta+\mathcal{L}_\xi$, the perturbed continuity equations take the form (see, e.g., \cite{friedman1978})
\begin{gather}
\label{Deltan}
\Delta n_k+n_{k0}\pnabla_\rho\xi^\rho=0.
\end{gather}
Multiplying this equation by $(\p p/\p n_k)_0$, by $\mu_{k0}$ or by $(\p\mu_m/\p n_k)_0$, and then performing the summation over the $k$ index, we obtain, with the use of thermodynamic relations, the three following equations:
\begin{gather}
\label{DeltaPeq}
\Delta p+\gamma p_0\pnabla_\rho \xi^\rho=0, \quad
\Delta\varepsilon+w_0\pnabla_\rho\xi^\rho=0, \\
\label{DeltaMuEq}
\Delta\mu_m+\biggl(\frac{\p p}{\p n_m}\biggr)_0\pnabla_\rho \xi^\rho=0,
\end{gather}
where the introduced coefficient $\gamma$ is related to the speed of sound $c_s$ in stellar matter with frozen composition as 
\begin{gather}
\gamma=\frac{w_0}{p_0}\biggl(\frac{c_s}{c}\biggr)^2, \quad c_s=c\sqrt{\frac{1}{w_0}\biggl(\frac{\p p}{\p n_k}\biggr)_0 n_{k0}}.
\end{gather}
From the derived equations it is easy to see, that the Lagrangain perturbations satisfy
\begin{gather}
\frac{\Delta n_k}{n_{k0}}=\frac{\Delta\mu_m}{(\p p/\p n_m)_0}=\frac{\Delta\varepsilon}{w_0}=\frac{\Delta p}{\gamma p_0}=\frac{\Delta w}{w_0+\gamma p_0}.
\end{gather}
From these relations it follows that, if one knows the Eulerian pressure perturbation and Lagrangian displacement, one may find Eulerian perturbations of the remaining thermodynamic quantities. In particular, for the energy density and enthalpy density perturbations one finds 
\begin{gather}
\label{energy}
\delta\varepsilon=\frac{w_0}{\gamma p_0}\delta p-w_0\xi^\rho \mathcal{A}_\rho, \\
\label{enthalpy}
\delta w=\biggl(1+\frac{w_0}{\gamma p_0}\biggr)\delta p-w_0\xi^\rho \mathcal{A}_\rho,
\end{gather}
where we have introduced 
\begin{gather}
\mathcal{A}_\rho\equiv\frac{1}{w_0}\nabla_\rho w_0-\frac{1}{w_0}\biggl(1+\frac{w_0}{\gamma p_0}\biggr)\nabla_\rho p_0 \equiv A(r)\delta_\rho^r.
\end{gather}
The function $A(r)$, sometimes referred to as the Schwarzschild discriminant, serves as a measure of the matter nonbarotropicity. In barotropic matter $A(r)=0$ and $\delta p=(d p/d \varepsilon)_0 \delta\varepsilon=(c_s/c)^2 \delta\varepsilon$, while in nonbarotropic one $A(r)\neq 0$. In the case of nonbarotropic stellar matter, we are interested in, $A(r)$ takes small (but nonzero!) values due to weak nonbarotropicity of stellar matter.


\subsection{Dissipative effects and gravitational radiation}\label{Sec II D}

The consideration in the previous sections does not account for the effects of dissipation and gravitational radiation. These effects cause the energy $E$ of a perturbed star to decrease with the rate $\dot{E}$ (do not confuse $E$ with the electric vector $E^\mu$). Let $\dot{E}_\zeta$ be the loss rate due to bulk viscosity, $\dot{E}_\eta$ -- due to shear viscosity, $\dot{E}_{\mathcal D}$ -- due to diffusion, and $\dot{E}_{\rm GW}$ -- due to gravitational radiation \footnote{We do not consider dissipation due to thermal conductivity since it is negligible compared to the viscous dissipation.}. Then the total energy loss rate, measured in the inertial reference frame, equals
\begin{gather}
\dot{E}=\dot{E}_{\zeta}+\dot{E}_{\eta}+\dot{E}_{\mathcal D}+\dot{E}_{\rm GW}.
\end{gather}

The energy loss rate due to dissipative mechanisms equals the heat generation rate in the course of a perturbation. From the viewpoint of a local observer, comoving with the fluid, a portion of heat $d\mathcal{Q}$, generated in the coordinate volume $d^3x=dr \ d\theta \ d\varphi$ with the temperature $T$ during the time $dt$, can be found from the equation \cite{tolman1987}
\begin{gather}
\frac{d\mathcal{Q}}{T}=(\nabla_\mu s^\mu)c \sqrt{-\rm g} \ dt \ d^3x, \quad {\rm g}\equiv\det{\rm g}_{\mu\nu},
\end{gather}
where $s^\mu$ is the entropy density four-current. Following \cite{tolman1987} we note that the right-hand side of the equation above is a relativistic invariant, so is the left-hand side, and, using the fact that the (redshifted) temperature measured by the distant observer equals $T^\infty=T/\Lambda$, we find the redshifted heat $dQ^\infty=dQ/\Lambda$, measured by the same observer:
\begin{gather}
d\mathcal{Q}^\infty=T^\infty (\nabla_\mu s^\mu)c \sqrt{-\rm g} \ dt \ d^3x.
\end{gather}
Adopting the general expression for the entropy generation rate $\nabla_\mu s^\mu$ from \cite{dg2021, dommes2020}, introducing supplementary tensors
\begin{gather}
\sigma^{\mu\nu}\equiv\pnabla^\mu \mathfrak{u}^\nu+\pnabla^\nu \mathfrak{u}^\mu-\frac{2}{3}\perp^{\mu\nu}(\pnabla_\lambda \mathfrak{u}^\lambda), \\
d_{k}^\mu\equiv\pnabla^\mu\biggl(\frac{\mu_{k}}{T}\biggr)-\frac{e_{k}E^\mu}{T},
\end{gather}
and integrating the heat generation rate $d\mathcal{Q}^\infty/dt$ over the stellar volume, we find
\begin{gather}
\label{EdotZeta}
\dot{E}_\zeta=-\int\zeta(\nabla_\mu \mathfrak{u}^\mu)^2 c^2\frac{\sqrt{-{\rm g}}}{\Lambda} \ d^3x, \\
\label{EdotEta}
\dot{E}_\eta=-\int\frac{\eta}{2}\sigma_{\mu\nu}\sigma^{\mu\nu} c^2 \frac{\sqrt{-{\rm g}}}{\Lambda} \ d^3x, \\
\label{EdotDraw}
\dot{E}_\mathcal{D}=-\int T \ \mathcal{D}_{km}d_{k}^\mu d_{m\mu} \ c \frac{\sqrt{-{\rm g}}}{\Lambda} \ d^3x,
\end{gather}
where with the accepted accuracy one has for the redshift factor: $1/\Lambda\approx e^\nu$. Bulk viscosity $\zeta>0$, shear viscosity $\eta>0$, and diffusion matrix $\mathcal{D}_{km}$ (symmetric, positive definite) are temperature-dependent kinetic coefficients, provided by the microscopic theory. One can show \cite{dommes2020,dg2021} that in degenerate matter diffusion coefficients satisfy $\mathcal{D}_{mk}\mu_{k}=0$. Also, due to the superconductivity of particle species ``$\rm s$'', diffusion coefficients related to ``${\rm s}$'' vanish, $\mathcal{D}_{{\rm s}k}=\mathcal{D}_{k{\rm s}}=0$ for any $k$ (including $k=\rm s$). It is easy to verify that in equilibrium $\dot{E}_\zeta=\dot{E}_\eta=0$, as expected. In order to ensure the absence of dissipation due to diffusion, the equilibrium chemical potentials of normal particle species must satisfy a condition: $d_{k0}^\mu=0$, which is consistent with \eqref{equilibriumT} and \eqref{equilibriumMU}. Note that the requirement $\dot{E}_\mathcal{D}=0$ does not lead to the condition $d_{{\rm s}0}^\mu=0$ for the superconducting component. The latter, nevertheless, holds and can be derived using the equilibrium superconducting equation, as discussed above.

As mentioned previously, gravitational radiation and dissipation only weakly affect stellar oscillations. For this reason the calculation of the energy change rates $\dot{E}_{\eta,\zeta,\mathcal{D},{\rm GW}}$ can be performed with the use of solutions of nondissipative hydrodynamic equations, discussed in the preceeding sections. Keeping this in mind, let us rewrite the expression for $\dot{E}_\mathcal{D}$ in a form, more convenient for further calculations. First, we exclude the self-consistent electric field via \eqref{Efield} and use the quasineutrality $e_k n_k=0$ and equilibrium condition \eqref{equilibriumMU2} with $m={\rm s}$ to obtain
\begin{gather}
\label{dk}
\begin{gathered}
d_{k\mu}=\frac{\Lambda}{T}\Phi_{k m}\pnabla_\mu\biggl[\frac{\delta\mu_{m}-e_{m}/e_{\rm s} \ \delta\mu_{\rm s}}{\Lambda}\biggr], \\
\Phi_{km}\equiv \delta_{km}-\frac{e_{k}}{e_{\rm s}}\biggl(\delta_{{\rm s}m}-\frac{\mu_{{\rm s}0}n_{m0}}{w_0}\biggr).
\end{gathered}
\end{gather}
Note that the supplementary matrix $\Phi_{km}$ satisfies $\Phi_{km}e_m=0$ and $\Phi_{km}\mu_{m0}=\mu_{k0}$. Then, using \eqref{DeltaPeq}, \eqref{DeltaMuEq} and \eqref{equilibriumMU2} with $m={\rm s}$, we find
\begin{multline}
\nabla_\mu\biggl[\frac{\delta\mu_m-e_{m}/e_{\rm s} \ \delta\mu_{\rm s}}{\Lambda}\biggr]=\\
=-\frac{\mu_{m0}-e_m/e_{{\rm s}0} \ \mu_{{\rm s}0}}{\Lambda}\nabla_\mu(\xi^\rho\nabla_\rho\ln\Lambda)+\\
+\nabla_\mu\biggl[\frac{1}{\Lambda}\biggl(\frac{\p p}{\p n_m}-\frac{e_m}{e_{\rm s}}\frac{\p p}{\p n_{\rm s}}\biggr)_0\frac{\Delta p}{\gamma p_0}\biggr].
\end{multline}
Finally, we substitute this result into \eqref{dk} and, using the mentioned properties of $\mathcal{D}_{km}$ and $\Phi_{km}$, obtain
\footnote{One can show that the resulting expression for $\dot{E}_\mathcal{D}$ also holds in the case when the temperature is perturbed in the course of oscillations, since the terms containing temperature gradients $\pnabla_\mu(T/\Lambda)$ vanish due to $\mathcal{D}_{km}\mu_m=0$.}
\begin{gather}
\label{EdotDiff}
\begin{gathered}
\dot{E}_\mathcal{D}=-\int  \frac{\Lambda^2}{T}  \ \hat{\mathcal{D}}_{pq}\hat{d}_{p}^\mu \hat{d}_{q\mu} \ c{e^\nu}\sqrt{-\rm g} \ d^3 x, \\
\hat{\mathcal{D}}_{pq}\equiv\mathcal{D}_{km}\Phi_{kp}\Phi_{mq}, \\
\hat{d}_{k\mu}\equiv \pnabla_\mu(\pi_{k}\Delta p), \quad \pi_{k}(r)=\frac{1}{\Lambda\gamma  p_0}\left(\frac{\partial p}{\partial n_{k}}\right)_0.
\end{gathered}
\end{gather}

To compute the energy loss rate due to the emission of gravitational waves, we use the general expressions, derived by Thorne \cite{thorne1980}. According to the formalism, developed therein, the energy loss rate of a neutron star associated with a periodic oscillation with frequency $\sigma$, equals
\begin{gather}
\label{EdotGW}
\begin{gathered}
\dot{E}_{\rm GW}=-\frac{G}{c^{3}}\sum_{L=2}^\infty \sum_{M=-L}^L \biggl(\frac{\sigma}{c}\biggr)^{2L+2}N_L(|\mathcal{I}_{LM}|^2+|\mathcal{S}_{LM}|^2), \\
N_L=\frac{4\pi}{[(2L+1)!!]^2}\frac{(L+1)(L+2)}{L(L-1)}, \raisetag{1cm}
\end{gathered}
\end{gather}
where $\mathcal{I}_{LM}$ and $\mathcal{S}_{LM}$ are the mass and current multipole moments, respectively. Once the effective stress-energy tensor of the perturbed neutron star is known, these quantities can be calculated using established formulas (see Appendix \ref{AppB} for technical details).

All the previously discussed expressions are written in the inertial reference frame, implying that $E$ includes the rotational energy of the star. Further, it is more convenient to use the corotating reference frame, where the energy perturbation $\tilde{E}$ coincides with the oscillation energy. In this frame the dissipative energy loss rates $\dot{\tilde{E}}_{\zeta,\eta,\mathcal{D}}=\dot{E}_{\zeta,\eta,\mathcal{D}}$ remain the same, while that due to gravitational radiation changes to $\dot{\tilde{E}}_{\rm GW}=(\sigma_r/\sigma)\dot{E}_{\rm GW}$ (see, e.g., \cite{ak2001}), where $\sigma_r$ is the oscillation frequency in the corotating reference frame. As a result, the total (oscillation) energy change rate in the corotating reference frame equals:
\begin{gather}
\dot{\tilde{E}}=\dot{E}_{\zeta}+\dot{E}_{\eta}+\dot{E}_{\mathcal D}+(\sigma_r/\sigma)\dot{E}_{\rm GW}.
\end{gather}
If, for a given perturbation, $\sigma_r$ and $\sigma$ are of opposite signs, the last term becomes positive, indicating that the perturbation is unstable with respect to the emission of gravitational waves (physically, the emission of gravitational waves triggers the transformation of the stellar rotation energy into the oscillation energy). This effect, known as the CFS instability, works against the energy leakage due to dissipative mechanisms. The condition $\dot{\tilde{E}}>0$ defines the instability window, i.e., those combinations of $(\Omega, T^\infty)$, for which the perturbation is unstable.

For periodic oscillations, each of the discussed mechanisms can be suitably characterized by the corresponding evolutionary timescale, defined as:
\begin{gather}
\label{timescales}
\frac{1}{\tau_{\zeta,\eta,\mathcal{D},{\rm GW}}}=\frac{1}{2\tilde{E}}\biggl|\langle \dot{\tilde{E}}_{\zeta,\eta,\mathcal{D},{\rm GW}} \rangle_P \biggr|,
\end{gather}
where $\langle\dots\rangle_P$ denotes the average over the oscillation period $P$. In the Cowling approximation, the energy $\tilde{E}$ can be found with the use of the perturbed conservation law, associated with the Killing vector $k^\rho=u^\rho/\Lambda$ (see Appendix \ref{AppA}):
\begin{multline}
\label{Etilde}
\tilde{E}=\int\biggl[\frac{w_0}{2}\delta\mathfrak{u}_\rho \delta\mathfrak{u}^\rho+\frac{1}{2}\frac{(\delta p)^2}{\gamma p_0}+\frac{1}{\Lambda}\delta p\delta\mathfrak{u}^t+\\
+\frac{1}{2}w_0 (\xi^\rho \mathcal{A}_\rho)(\xi^\lambda\nabla_\lambda \ln\Lambda)\biggr]\sqrt{-{\rm g}} \ d^3 x.
\end{multline}
%


\section{r-mode instability windows}


\subsection{Nonanalytic r-modes in a nutshell}

In this study, we, for simplicity, ignore the oblateness of the rotating stellar model [i.e., terms $O(\Omega^2)$ in \eqref{geometry} and \eqref{Lambda}] and formally treat the framedragging effect as weak, i.e. we use the equations, derived under the assumption that the metric function $\omega(r)=\epsilon\tilde{\omega}(r)$ is small, $\epsilon=\max\omega(r)\ll 1$. Although in reality framedragging effect cannot be considered as weak throughout the entire star, it significantly weakens as one approaches the stellar surface (for the NS model, employed in this study, we have $\omega\sim 0.1$ near the surface; more details on the NS model will be given in Sec.\,\ref{III B}). At the same time, as we shall see, at slow rotation rates $r$-modes tend to localize in the vicinity of the stellar surface, where the framedragging effect is, indeed, weak. As a result, the slower the star rotates, the more accurate the weak framedragging approximation. Although for larger rotation rates the weak framedragging approximation may be not too accurate, we believe that it still provides qualitatively correct relativistic $r$-mode description (see Refs.\ \cite{kgk2022_1, kgk2022_2} for a more detailed discussion of this issue).

Recently we have shown that, under these assumptions, the general system \eqref{hydrodynamics} admits the solution in the form of nonanalytic $r$-modes -- predominantly toroidal oscillations with discrete eigenfrequency spectrum and eigenfunctions, demonstrating nonanalytic behavior in $\Omega$ for extremely slow rotation rates. Below, we only briefly summarize the properties of these $r$-modes, and refer the interested reader to Refs.\ \cite{kgk2022_1, kgk2022_2} for details.

Various $r$-mode solutions differ by quantum numbers $(l,m)$, which determine the angular dependence of the leading contribution to the toroidal function ${\rm T}$ in the slow rotation limit: ${\rm T}\sim P_{l}^m(x)e^{im\varphi}$, where $x=\cos\theta$ and $P_{l}^m(x)$ is the associated Legendre polynomial. From the viewpoint of CFS instability, the most interesting $r$-mode is the one that contributes the most to the $L=2$ terms in $\dot{E}_{\rm GW}$. Below we restrict ourselves to the $l=m=2$ mode, which is currently believed to be the most CFS unstable, at least in Newtonian theory (see, however, comments below). The eigenfunctions and oscillation frequency of this mode possess the following structure:
\begin{gather}
\label{r-mode structure}
\begin{gathered}
\sigma=\Omega[\sigma^{(0)}+\sigma^{(1)}], \\
\mathrm{T}=-i[T_m(r)P_m^m(x)+T_{m+2}(r)P_{m+2}^m(x)]e^{i\sigma t+im\varphi}, \\
\mathrm{Q}=Q_{m+1}(r)P_{m+1}^m(x)e^{i\sigma t+im\varphi}, \\
\xi^r=\xi_{m+1}(r)P_{m+1}^m(x)e^{i\sigma t+im\varphi}, \\
\delta p=\delta p_{m+1}(r)P_{m+1}^m(x)e^{i\sigma t+im\varphi}, \\
\delta w=\delta w_{m+1}(r)P_{m+1}^m(x)e^{i\sigma t+im\varphi}, \\
\delta \varepsilon=\delta \varepsilon_{m+1}(r) P_{m+1}^m(x)e^{i\sigma t+im\varphi},
\end{gathered}\raisetag{1.5cm}
\end{gather}
where the leading frequency contribution equals
\begin{gather}
\label{sigma0}
\sigma^{(0)}=\sigma_r^{(0)}-m=\frac{2m}{m(m+1)}-m.
\end{gather}
Eigenfrequency correction $\sigma^{(1)}$ and all the eigenfunctions except $T_m$ are small due to weak framedragging effect, slow stellar rotation, or both (see appendix \ref{AppD} for the comment on the possible violation of the $r$-mode ordering due to the interplay between weak matter nonbarotropicity and rapid stellar rotation). 

The eigenfrequency correction $\sigma^{(1)}$, toroidal function $T_m(r)$ and radial displacement $\xi_{m+1}(r)$ are found from the  system [see Appendix \ref{AppC} for the explicit form of the coefficients $C_{1,2,3}(r)$ and $G_{1,2}(r)$]:
\begin{gather}
\label{r-modes sys}
\left\{
\begin{gathered}
\biggl[C_{1}(r)\frac{d}{d r}+C_{2}(r)\biggr]\xi_{m+1}+ \hfill \\
\hfill +\biggl[\Omega^2 C_{3}(r)+\sigma^{(1)}+\frac{2\epsilon\tilde{\omega}(r)}{m+1}\biggr]T_m=0 \\
\biggl[\frac{d}{dr}+G_1(r)\biggr]T_m+\frac{G_2(r)}{\Omega^2}\xi_{m+1}=0. \hspace{1.2cm}
\end{gathered}
\right.
\end{gather}
The sought eigenfunctions are required to 1) be regular at the stellar center and 2) correspond to the vanishing total pressure at the stellar surface. Once the system is solved, the remaining eigenfunctions can be found using \eqref{energy}, \eqref{enthalpy}, \eqref{r-modes sys} and relations [see Appendix \ref{AppC} for $\Pi(r)$, $q_{1,2,3}(r)$, and $t_{1,2,3}(r)$]
\begin{gather}
\label{eigenfunctions}
\begin{gathered}
\delta p_{m+1}(r)=\Omega^2\Pi(r)T_m, \\
Q_{m+1}(r)=\biggl\{q_1\biggl[\sigma^{(1)}+\frac{2\epsilon\tilde{\omega}(r)}{m+1}\biggr]+\Omega^2 q_2(r)\biggr\}T_m+ \hfill \\
\hfill +\Omega^2 q_3(r) T_m', \\
T_{m+2}(r)=\biggl\{t_1\biggl[\sigma^{(1)}+\frac{2\epsilon\tilde{\omega}(r)}{m+1}\biggr]+\Omega^2 t_2(r)\biggr\}T_m+ \hfill \\
\hfill +\Omega^2 t_3(r) T_m',
\end{gathered}
\end{gather}
where the prime denotes the derivative $d/dr$. The resulting solutions are classified by the number of nodes of $T_m(r)$. Nodeless $r$-mode is referred to as the fundamental $r$-mode in the literature, and is believed to be the most CFS unstable.

Depending (mainly) on the values of $\Omega^2$ and $\epsilon$, we can formally distinguish the three possibilities for the system \eqref{r-modes sys}:

\begin{enumerate}
\item[1)] {\it Newtonian limit.}
In order to make transition to the Newtonian $r$-mode equations we first, keeping in mind that $g(r)=\nu'(r)=-p_0'(r)/w_0(r)\sim (1/c^2)$, retain the leading in $(1/c^2)$ contributions to the coefficients, appearing in equations (\ref{r-modes sys}), (\ref{eigenfunctions}). Then we set all the metric functions, including the framedrag $\epsilon\tilde{\omega}(r)$, equal to zero
\footnote{At that we still retain the function $g(r)$.}.
The resulting equations coincide with the $r$-mode equations in the Newtonian theory (see, e.g., \cite{provost1981}). The corresponding $r$-mode eigenfunctions are analytic functions of $\Omega$, scaling as
\begin{gather}
\label{NewtOrdering}
\begin{gathered}
\sigma^{(1)}_{\rm Newt}\propto\Omega^2, \\
\xi_{m+1}^{\rm Newt}\sim Q_{m+1}^{\rm Newt} \sim T_{m+2}^{\rm Newt} \sim\Omega^2 T_{m}^{\rm Newt}, \\
\delta p_{m+1}^{\rm Newt} \sim\delta w_{m+1}^{\rm Newt}\sim \delta\varepsilon_{m+1}^{\rm Newt}\sim \Omega^2 T_{m}^{\rm Newt}.
\end{gathered}
\end{gather}
In what follows, we will refer to these and similar expressions, which demonstrate the relations between the $r$-mode eigenfunctions, as {\it the $r$-mode ordering}. It is worth noting that the ordering \eqref{NewtOrdering} would also be valid for relativistic $r$-modes, if one would completely ignore the framedragging effect.


\begin{figure}
\includegraphics[width=1.0\linewidth]{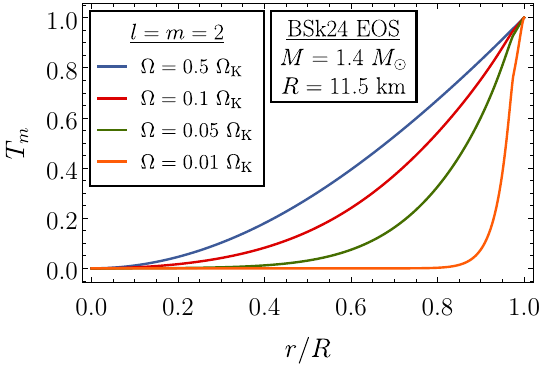}
\vspace{-0.8cm}
\caption{Toroidal function of the fundamental relativistic $l=m=2$ $r$-mode, calculated for different stellar rotation rates using BSk24 EOS for a star with the mass $M=1.4 \ M_\odot$ (where $M_\odot$ is the solar mass) and radius $R=11.5 \ {\rm km}$. The normalization condition $T_m(R)=1$ is implied. Details of the employed stellar model are discussed in Sec.\ref{III C}.}
\label{nonanalytic_mode}
\end{figure}


\vspace{1pt}

\item[2)] {\it Relativistic $r$-modes with $\Omega^2\gg \epsilon$.}
In this limit, the terms, containing the framedrag $\epsilon\tilde{\omega}(r)$, are negligibly small, so that relativistic $r$-modes effectively obey the Newtonian ordering \eqref{NewtOrdering}. In our numerical calculations (see Sec.\ \ref{III C} for the details of the employed NS model) this limit is formally achieved at $\Omega \gtrsim 1.5 \ \Omega_{\rm K}$
\footnote{We would like to stress that here we treat $\Omega$ and $\epsilon$ as {\it formal} parameters of the system \eqref{r-modes sys}. Of course, stars with such high rotation rates do not exist and in the case of the NS model, considered in the present study, the limit $\Omega^2\gg \epsilon$ is physically unreachable.}.
As $\Omega$ decreases towards the slower realistic rotation rates, the properties of relativistic $r$-modes start to differ from those, predicted by \eqref{NewtOrdering}.
\item[3)]  {\it Relativistic $r$-modes with $\Omega^2\ll \epsilon$.}
In this limit, the difference of the relativistic $r$-modes from their Newtonian cousins becomes especially pronounced. In this case, the analysis of the system \eqref{r-modes sys} reveals \cite{kgk2022_1, kgk2022_2} that $\sigma^{(1)}\sim\epsilon$, $\xi_{m+1}\sim\sqrt{\epsilon}\Omega T_m$ and that its solution is described by nonanalytic functions of $\Omega$, such that the operator $d/dr$ produces extra $\sqrt{\epsilon}/\Omega$ when acting on them, i.e.,  $d/dr\sim\sqrt{\epsilon}/\Omega$ (as an example, one may think of the function $e^{\sqrt{\epsilon}r/\Omega}$). As a result, in this limit, the relativistic $r$-mode ordering significantly differs from the Newtonian one \eqref{NewtOrdering}:
\begin{gather}
\label{ordering}
\begin{gathered}
\sigma^{(1)}\sim \epsilon, \quad Q_{m+1}\sim T_{m+2}\sim \epsilon T_m, \\
\xi_{m+1}\sim\delta w_{m+1}\sim\delta \varepsilon_{m+1}\sim \epsilon\kappa T_m, \\
\delta p_{m+1}\sim\epsilon\kappa^2 T_m, \ d/dr \sim 1/\kappa,
\end{gathered}
\end{gather}
where we have introduced the parameter of nonanalyticity $\kappa\equiv \Omega/\sqrt{\epsilon}$. Another important feature of relativistic $r$-modes in this limit is that, due to nonanalyticity, toroidal eigenfunction (and, therefore, all other eigenfunctions) nontrivially depends on $\Omega$, in contrast to its $\Omega$-independent Newtonian analogue (see Fig.~\ref{nonanalytic_mode} for illustration). This dependence becomes increasingly evident as $\Omega$ decreases, ultimately resulting in the localization of the modes near the stellar surface in the limit of extremely slow rotation rates. Such localization is naturally accompanied by the steepening of the gradients of the $r$-mode eigenfunctions in the vicinity of the stellar surface.
\end{enumerate}

\vspace{1pt}

Thus, we expect the typical timescales $\tau_{\zeta,\eta,\mathcal{D},{\rm GW}}$ to drastically differ from their Newtonian counterparts, at least for slow stellar rotation. In fact, as we shall see, the peculiar behavior of the relativistic $r$-modes leads to the significant amplification of the energy dissipation due to bulk viscosity and diffusion even at high rotation rates, where the effect of the mode localization is weak.

Another intriguing feature of relativistic $r$-modes is that the mass multipole contribution to gravitational radiation from the $(l,m)=(3,2)$ $r$-mode appears to be of the same order in $\Omega$ as the current multipole contribution from the discussed above $(l,m)=(2,2)$ $r$-mode. As a result, it is not completely clear, which of these modes is more unstable. Comparing the corresponding instability windows is an intriguing problem that deserves attention in future studies. In this work we, as announced, consider only the $l=m=2$ $r$-mode.


\subsection{r-mode energy and energy change rates}\label{III B}

In this section, we provide the approximate expressions for the $r$-mode energy and previously discussed energy change rates, valid in the limit of slow stellar rotation. Their derivation for both relativistic and Newtonian cases is lengthy, but rather straightforward and, basically, consists of the following steps:

\vspace{4pt}

\begin{enumerate}

\item[1)]
the solution in the form \eqref{r-mode structure} is substituted into the equations \eqref{EdotZeta}, \eqref{EdotEta}, \eqref{EdotDiff}, \eqref{EdotGW} and \eqref{Etilde};

\vspace{2pt}

\item[2)]
one uses equations \eqref{energy}, \eqref{enthalpy}, \eqref{r-modes sys} and \eqref{eigenfunctions} to write the result in terms of $T_m(r)$ and $T_m'(r)$;

\vspace{2pt}

\item[3)] one retains the terms up to the desired orders in $\Omega$ and $\epsilon$;

\vspace{2pt}

\item[4)]
one performs the integration of the obtained expressions over the $\theta$ and $\varphi$ angles.

\end{enumerate}

\vspace{4pt}

\noindent In the derivation, one should remember that the previously described $r$-mode solution is given in a complex form, while the energy and energy change rates, being second order quantities in Eulerian perturbations, are calculated with the physical real-valued $r$-mode solution. For any complex eigenfunction $f$ we define the real-valued physical eigenfunction $f_{\rm phys}$ as the real part of $f$:
\begin{gather}
f_{\rm phys}=(1/2)(f+f^\star).
\end{gather}
It is easy to show that, with such notation, the average of the product $(f_{1,{\rm phys}}f_{2,{\rm phys}})$ over the oscillation period equals
\begin{gather}
\label{fPhysAverage}
\langle f_{1,{\rm phys}} \ f_{2,{\rm phys}}\rangle_P=\frac{1}{4}(f_1 f_2^\star+f_1^\star f_2).
\end{gather}
In particular, we have $\langle f_{\rm phys}^2\rangle_P=(1/2)f f^\star$, where the factor $1/2$ can  naturally be interpreted as a result of averaging of $\sin^2(t/P)$ or $\cos^2(t/P)$ over the oscillation period.

The formula \eqref{fPhysAverage} allows one to calculate the $r$-mode energy \eqref{Etilde} and the averaged energy loss rates \eqref{EdotZeta}, \eqref{EdotEta}, \eqref{EdotDiff}, and \eqref{EdotGW} using the complex $r$-mode eigenfunctions, discussed in the previous section. We would also like to note that the calculation of the mass and current multipole moments in terms of the physical $r$-mode solution can be replaced with that in terms of the complex solution according to the following relations (note that in derivation of these relations we explicitly assumed $m\neq 0$):
\begin{gather}
\begin{gathered}
|\mathcal{I}_{L,M}[f_{\rm phys}]|=|\mathcal{I}_{L,-M}[f_{\rm phys}]|=\frac{\delta_{M,m}+\delta_{M,-m}}{2}|\mathcal{I}_{L,m}[f]|, \\
|\mathcal{S}_{L,M}[f_{\rm phys}]|=|\mathcal{S}_{L,-M}[f_{\rm phys}]|=\frac{\delta_{M,m}+\delta_{M,-m}}{2}|\mathcal{S}_{L,m}[f]|.
\end{gathered}
\end{gather}
As a result, the formula \eqref{EdotGW} takes the form
\begin{gather}
\dot{E}_{\rm GW}=-\frac{1}{2}\frac{G}{c^{3}}\sum_{L=2}^\infty\biggl(\frac{\sigma}{c}\biggr)^{2L+2}N_L(|\mathcal{I}_{Lm}[f]|^2+|\mathcal{S}_{Lm}[f]|^2).
\end{gather}

Using the formulas, described above, we derived the relativistic expressions for the $r$-mode energy and energy change rates, valid for slow rotation rates irrespectively of the relation between $\Omega^2$ and $\epsilon$ (in particular, they reduce to their Newtonian counterparts in the Newtonian limit). Some of them can be simplified by introducing the supplementary functions $h_{1,2}(r)$ as
\begin{gather}
\label{h1func}
T_m(r)=r^m e^{-(m-1)\nu}h_1(r), \\
h_2(r)=\frac{r^{m+1}w_0  e^{-(m+1)\nu }}{A}\frac{d h_1}{dr}.
\end{gather}
The function $h_1$, as we shall see, will play an important role in the further analysis. For convenience, we also introduce the following supplementary combinations of diffusion coefficients:
\begin{gather}
\begin{gathered}
\hat{\mathcal{D}}_1(r)=\pi_{k}\hat{\mathcal{D}}_{km}\pi_{m}, \quad \hat{\mathcal{D}}_2(r)=r\frac{d\pi_{k}}{dr}\hat{\mathcal{D}}_{km}\pi_{m}, \\
\hat{\mathcal{D}}_3(r)=r^2\frac{d\pi_{k}}{dr}\hat{\mathcal{D}}_{km}\frac{d\pi_{m}}{dr}, \\
\hat{\mathcal{D}}_4(r)=e^{2\lambda}(m+1)(m+2)\hat{\mathcal{D}}_1(r)+\hat{\mathcal{D}}_3(r),
\end{gathered}
\end{gather}
In these notations, the obtained expressions can be written as [$\alpha$ is the (conveniently chosen) normalization constant, related to the $r$-mode amplitude; note that $\alpha$ depends on the azimuthal quantum number $m$]
\begin{widetext}
\begin{gather}
\label{ETildeRmode}
\tilde{E}=\frac{\alpha^2\Omega^2}{2 c^2}\int\limits_0^R w_0 T_m^2 r^2 e^{\lambda-\nu} \ dr, \\
\label{EdotZetaRmode}
\langle\dot{\tilde{E}}_\zeta\rangle_P=-\frac{16 m \alpha^2 \Omega ^6}{c^4 (m+1)^5 (2 m+3)}\int\limits_0^R \frac{\zeta  e^{\lambda-(2 m+3) \nu}}{A^2}\left(\frac{c}{c_s}\right)^4\biggl(\frac{d h_1}{dr}\biggr)^2 r^{2 m+4} { e^\nu} \ dr,\\
\label{EdotEtaRmode}
\langle\dot{\tilde{E}}_\eta\rangle_P=-\alpha^2\Omega ^2\int\limits_0^R \eta \biggl[\left(T_m-r \frac{d}{dr}T_m\right)^2+(m-1) (m+2) e^{2 \lambda} T_m^2\biggr]e^{-\lambda -\nu } \ { e^\nu} dr, \\
\label{EdotDRmode}
\langle\dot{\tilde{E}}_\mathcal{D}\rangle_P=-\frac{4 m \alpha^2\Omega ^4}{c^3 (m+1)^3 (2 m+3)}\int\limits_0^R \frac{\Lambda^2}{T}\biggl[ r^2 \biggl(\frac{d h_2}{d r}\biggr)^2 \hat{\mathcal{D}}_1+2 r h_2\frac{d h_2}{dr} \hat{\mathcal{D}}_2+ h_2^2 \hat{\mathcal{D}}_4 \biggr]e^{\nu-\lambda} \ { e^\nu} dr, \\
\label{EdotGWRmode}
\langle\dot{\tilde{E}}_{\rm GW}\rangle_P=32\pi G \alpha^2 \frac{2^{2 m} (m-1)^{2 m}}{(m+1)^4}\left[\frac{(m+2)!}{(2 m+1)!}\right]^2\left(\frac{m+2}{m+1}\right)^{2 m}\frac{\Omega ^2}{c}\left(\frac{\Omega }{c}\right)^{2 m+2}\biggl[\frac{1}{c^2}\int\limits_0^R  w_0 T_m e^{2 \lambda} r^{m+2} dr \biggr]^2,
\end{gather}
\end{widetext}

In order to make the transition to the Newtonian limit in the equations above, we set all the metric functions to zero, including those in the definitions of $h_1$, $h_2$, and $\hat{\mathcal{D}}_4$. Note that, although in our model $\Lambda=e^{-\nu}$, we retain the redshift $\Lambda$ along with its derivatives. This approach corresponds to the consideration of the Newtonian oscillations on the relativistic background, i.e., accounting for the relativistic equilibrium distribution of the temperature \eqref{equilibriumT} and chemical potentials \eqref{equilibriumMU2} inside the star. We also stress that, while for Newtonian $r$-modes all the terms in the expressions above are of the same order in $\Omega$, for relativistic $r$-modes some of them turn out to be of lower order in $\Omega$ than the others because of the more complicated relativistic ordering.

We see that, at slow rotation rates, due to the $r$-mode nonanalyticity ($d/dr\sim 1/\kappa$), the subintegral expressions in $\langle \dot{\tilde{E}}_{\zeta,\eta} \rangle_P$ are effectively amplified by a factor $1/\Omega^2$ (more accurately, by $1/\kappa^2$), while that in $\langle \dot{\tilde{E}}_\mathcal{D}\rangle_P$ is effectively amplified even more strongly -- by a factor $1/\Omega^4$ (more accurately, by $1/\kappa^4$)! Moreover, one should also account for the fact that at slow rotation rates nonanalytic $r$-modes are localized in a tiny region in the vicinity of the stellar surface, whose size itself depends on the angular velocity and tends to zero at $\Omega\to 0$ \cite{kgk2022_1, kgk2022_2}. As a result, the dependence of expressions \eqref{ETildeRmode}--\eqref{EdotGWRmode} [and, therefore, the evolutionary timescales $\tau_{\zeta,\eta,\mathcal{D},{\rm GW}}$] on $\Omega$ turns out to be significantly more complicated than in the Newtonian theory and, generally, cannot be explicitly determined, unless one considers the limit of extremely slow rotation. In this limit, we discard the relatively small terms in subintegral expressions, and, using the system \eqref{r-modes sys} in order to express the derivative $T_m''(r)$ through $T_m(r)$, find that
\begin{widetext}
\begin{gather}
\label{Etilde0}
\tilde{E}\approx\frac{\alpha^2\Omega^2}{2 c^2}\int\limits_0^R w_0 T_m^2 e^{\lambda-\nu}r^2 dr, \\
\label{EdotVisc0}
\langle\dot{\tilde{E}}_\zeta\rangle_P\approx-\frac{16 m \alpha^2 \Omega^6}{(m+1)^5(2m+3)}\int\limits_0^R \frac{\zeta T_m'{}^2}{A^2 c_s^4} r^4 e^{\lambda-5 \nu} { e^\nu} dr, \\
\langle\dot{\tilde{E}}_\eta\rangle_P\approx-\alpha^2\Omega ^2\int\limits_0^R \eta \ T_m'{}^2 e^{-\lambda-\nu} r^2 { e^\nu} dr, \\
\langle\dot{\tilde{E}}_\mathcal{D}\rangle_P\approx-\frac{\alpha^2 c (2m+3)(m+1)^3}{16 m}\int\limits_0^R \frac{g^2 w_0^2}{T}[(m+1)\sigma^{(1)}+2 \epsilon  \tilde{\omega}(r)]^2 \ \hat{\mathcal{D}}_1(r) \ T_m^2 e^{-\lambda-\nu} { e^\nu} dr, \\
\label{EdotGW0}
\langle\dot{\tilde{E}}_{\rm GW}\rangle_P\approx32\pi G \alpha^2 \frac{2^{2 m} (m-1)^{2 m}}{(m+1)^4}\left[\frac{(m+2)!}{(2 m+1)!}\right]^2\left(\frac{m+2}{m+1}\right)^{2 m}\frac{\Omega ^2}{c}\left(\frac{\Omega }{c}\right)^{2 m+2}\biggl[\frac{1}{c^2}\int\limits_0^R  w_0 T_m e^{2 \lambda} r^{m+2} dr \biggr]^2.
\end{gather}
\end{widetext}
Further, at $\Omega\to 0$, the $r$-mode is localized in the region $r_t\leq r\leq R$ and exponentially suppressed at $0\leq r<r_t$ \cite{kgk2022_2}, where
\begin{gather}
r_t=R+z_0\biggl(\frac{\kappa}{\beta}\biggr)^{2/3}, \quad \beta^2=-\frac{2\tilde{\omega}'(R)}{m+1}\frac{G_2(R)}{C_1(R)},
\end{gather}
and $z_0$ is the (closest to zero) root of the derivative of the first type Airy function: $\Ai'(z_0)=0$. Therefore, in the formulas above, one can use the approximate expression for the toroidal function, which is accurate in the region of the mode localization and near the point $r_t$ (the normalization constant can be safely absorbed
in $\alpha$) \cite{kgk2022_2}:
\begin{gather}
T_m(r)\approx \Ai[z(r)], \quad z(r)=(r_t-r)\biggl(\frac{\beta}{\kappa}\biggr)^{2/3}. 
\end{gather}
Although for $0\leq r<r_t$, far from the point $r_t$, this formula is not too accurate, the function $\Ai[z(r)]$ just like the actual toroidal function $T_m(r)$ is exponentially suppressed in this region, so that the replacement $T_m(r)\to\Ai[z(r)]$ in the whole area of integration cannot lead to significant errors. Then the problem reduces to the calculation of the integrals of the form
\begin{gather}
I_{kn}[f]=\int\limits_{0}^R f(r) \biggl(\frac{d^{k}}{dr^k} \Ai[z(r)]\biggr)^n dr,
\end{gather}
where $f(r)$ is some $\kappa$-independent function of $r$. Changing the integration variable from $r$ to $z(r)$, we obtain
\begin{multline}
I_{kn}[f]=(-1)^{kn}\biggl(\frac{\beta}{\kappa}\biggr)^{(2/3)(kn-1)}\times \\
\times\int\limits_{z_0}^{r_t(\beta/\kappa)^{2/3}}f\biggl[R+(z_0-z)\biggl(\frac{\kappa}{\beta}\biggr)^{2/3}\biggr]\biggl[\frac{d^k}{dz^k} \Ai(z)\biggr]^n dz. \raisetag{2cm}
\end{multline}
In the $\Omega\to 0$ limit the upper limit in the integral can be replaced to infinity, and function $f$ can be replaced by the leading term of its Taylor series near the stellar surface: $f(r)\approx f_R (r-R)^{n_f}$. As a result we have
\begin{gather}
I_{kn}[f]\approx (-1)^{kn}f_R\biggl(\frac{\beta}{\kappa}\biggr)^{(2/3)(kn-n_f-1)} J_{knn_f}, \\
J_{knn_f}\equiv\int\limits_{z_0}^\infty (z_0-z)^{n_f}\biggl[\frac{d^k}{dz^k}\Ai(z)\biggr]^n dz.
\end{gather}
We see that three integer numbers $n$, $k$ and $n_f$ define the dependence of the considered integrals on $\kappa$ and, therefore, on $\Omega$. Let us, for example, estimate the dependence of $\tau_{\rm GW}$ on $\Omega$. In our model, for the corresponding integrals $k=0$ and $n_f=1$. Then we find that $\tilde{E}\propto \Omega^2 \Omega^{4/3}$ and $\langle\dot{\tilde{E}}_{\rm GW}\rangle_P\propto \Omega^2 \Omega^{2m+2} (\Omega^{4/3})^2$, leading to $\tau_{\rm GW}\propto 1/\Omega^{2m+2+4/3}$, which differs from the traditional result $\tau\propto 1/\Omega^{2m+2}$, valid for Newtonian $r$-modes (a result we have mentioned in \cite{kgk2022_2}). Similar estimates can also be made for $\tau_{\zeta,\eta,\mathcal{D}}$.


\subsection{Evolutionary timescales and instability windows}\label{III C}

In this section, we provide the results of our numerical calculation of the evolutionary timescales $\tau_{\eta,\zeta,\mathcal{D},{\rm GW}}$ \eqref{timescales} and instability windows for relativistic and Newtonian $r$-modes, using \eqref{ETildeRmode}-\eqref{EdotGWRmode}. We remind the reader that for simplicity in this study we, unless stated otherwise, ignore the neutron star crust and consider a toy model of a neutron star, consisting only of the liquid core (see Sec. \ref{IV Conclusion} for the discussion of the crust-associated effects).


\begin{figure*}
\centering\includegraphics[width=0.87\linewidth]{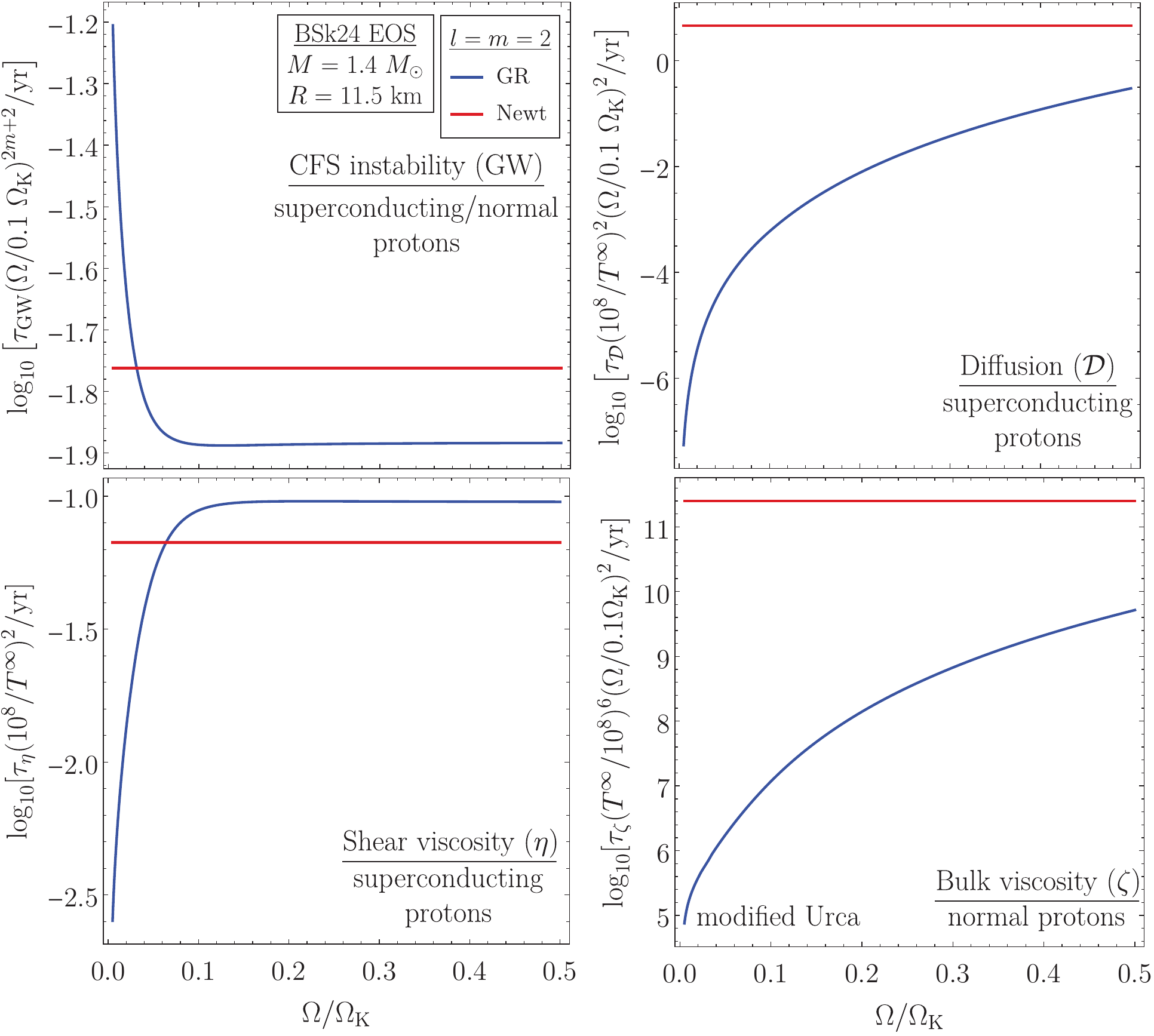}
\vspace{-0.4cm}
\caption{Relativistic (``GR'', blue lines) and Newtonian (``Newt'', red lines) $l=m=2$ fundamental $r$-mode evolutionary timescales, associated with the CFS instability (upper left panel), diffusion (upper right panel), shear viscosity (lower left panel) and bulk viscosity (due to modified Urca processes; lower right panel).}
\label{timescalesplt}
\end{figure*}



\begin{figure}[H]
\includegraphics[width=1.0\linewidth]{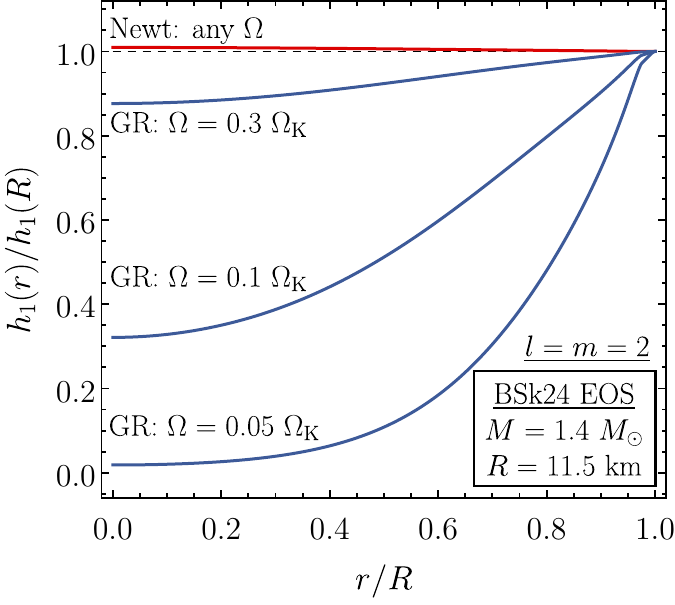}
\vspace{-0.6cm}
\caption{The ratio $h_1(r)/h_1(R)$ in the Newtonian theory (red line; does not depend on $\Omega$) and in GR (blue lines).}
\label{h1plot}
\end{figure}



\begin{figure}[H]
\includegraphics[width=1.0\linewidth]{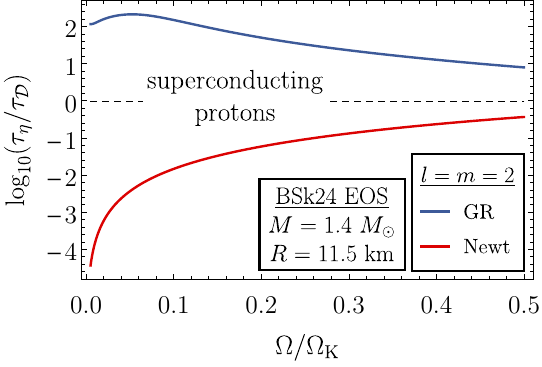}
\vspace{-0.7cm}
\caption{Comparison of the $l=m=2$ fundamental $r$-mode damping timescales due to shear viscosity and particle diffusion in general relativity (``GR'', blue line) and Newtonian theory (``Newt'', red line).}
\label{timescalesGRvsNewt}
\end{figure}


As a microphysical input, we use a neutron star with the mass $M=1.4 \ M_\odot$ and radius $R=11.5 \ {\rm km}$, described by the BSk24 EOS \cite{gorielyetal2013}
\footnote{Note that this model differs from that adopted in Ref.\ \cite{kgk2022_1}. It is obtained by stopping the integration of the Tolman–Oppenheimer–Volkoff equations at the density, corresponding to the crust-core interface of the model, adopted in Ref.\ \cite{kgk2022_1} (the pressure in this point equals approximately $5\times 10^{-3} \ p_{\rm c}$, where $p_{\rm c}$ is the central pressure). Thus, the surface of the NS model in the present study formally corresponds to the the crust-core interface of the model in Ref.\ \cite{kgk2022_1}. In fact, this is the model that we used in Ref.\ \cite{kgk2022_2}.}.
Its matter consists of neutrons $(n)$, protons $(p)$, electrons $(e)$ and muons $(\mu)$. To calculate $\tau_{\eta,\mathcal{D}}$ we assume that protons are strongly superconducting and adopt the shear viscosity $\eta$ from \cite{shternin2018} and diffusion matrix $\mathcal{D}_{km}$ from \cite{dommes2020}. Generally, to account for superconductivity in $\mathcal{D}_{km}$ one should replace the number density $n_p$ with $n_{p}^{\rm ex}$, the number density of proton Bogoliubov excitations (``normal'' proton component).
In our case of strong proton superconductivity $n_{p}^{\rm ex}=0$, which effectively reduces the number of particle species by one (i.e., in the derivation of $\mathcal{D}_{km}$ the 4-component $npe\mu$-matter is effectively considered as 3-component $ne\mu$-matter). As for the energy dissipation due to bulk viscosity, the latter is, strictly speaking, strongly suppressed by the proton superconductivity and is unlikely to play a significant role in our model. Nevertheless, keeping in mind future application of the theory to the hyperonic stellar matter, where the bulk viscosity is much stronger and might be important \cite{ofengeimetal2019}, it is still interesting to look how the dissipation due to bulk viscosity is affected by the peculiar properties of relativistic $r$-modes. To address this question, we will consider a simplified problem by ignoring the proton superconductivity when calculating $\zeta$ generated (in the $npe\mu$-matter) by the modified Urca processes \cite{haenseletal2001} [note that $\zeta$ depends on the {\it local} oscillation frequency $\sigma_{\rm loc}=(\sigma+m\Omega)\Lambda$]. Even for superconducting $npe\mu$-matter this approach is not completely unjustified, since, as we shall see, the bulk viscosity appears to be efficient only at very high temperatures, where not all protons are superconducting.

Figure \ref{timescalesplt} shows the relativistic (``GR'', blue lines) and Newtonian (``Newt'', red lines) fundamental $l=m=2$ $r$-mode evolutionary timescales. In the upper
left panel we see that at $\Omega\ll\Omega_{\rm K}$ $\tau_{{\rm GW}, {\rm GR}}\gg \tau_{{\rm GW}, {\rm Newt}}$, while at realistic rotation rates $\tau_{{\rm GW}, {\rm GR}}\sim\tau_{{\rm GW}, {\rm Newt}}$. Such behavior of relativistic $r$-modes is a result of the mode localization in the $\Omega\to 0$ limit, leading to $\tau_{{\rm GW}, {\rm GR}}/\tau_{{\rm GW}, {\rm Newt}}\sim \Omega^{-4/3}$. Lower left panel shows that at realistic rotation rates $\tau_{\eta, {\rm GR}}\sim\tau_{\eta, {\rm Newt}}$, while at extremely slow rotation rates $\tau_{\eta, {\rm GR}}\ll \tau_{\eta, {\rm Newt}}$, in contrast to the case of $\tau_{\rm GW}$. This behavior also arises due to the $r$-mode localization and is related to the fact that the subintegral expression \eqref{EdotEtaRmode} for $\langle \dot{\tilde{E}}_\eta\rangle_P$, unlike the equation \eqref{EdotGWRmode} for $\langle \dot{\tilde{E}}_{\rm GW}\rangle_P$, contains $T_m'{}^2(r)$, which is extremely large in the $\Omega\to 0$ limit, since $d/dr\sim 1/\Omega$. Thus, despite the mode being trapped in a narrow region near the stellar surface, its steep gradient eventually leads to $\tau_{\eta, {\rm GR}}\ll \tau_{\eta, {\rm Newt}}$.

Finally, the most interesting results are obtained for $\tau_\mathcal{D}$ (upper right panel) and $\tau_\zeta$ (lower right panel). Like for $\tau_{\eta, {\rm GR}}$, the presence of the derivatives of $T_m(r)$ in \eqref{EdotZetaRmode} and \eqref{EdotDRmode} significantly reduces $\tau_{\mathcal{D}, {\rm GR}}$ and $\tau_{\zeta, {\rm GR}}$ compared to their Newtonian values. This, however, does not explain the reduction of $\tau_{\zeta,{\rm GR}}$ and $\tau_{\mathcal{D}, {\rm GR}}$ at high rotation rates, $\Omega\gtrsim 0.1 \ \Omega_{\rm K}$. To explain this reduction we note that in the Newtonian theory the function $h_{1,{\rm Newt}}=r^{-m}T_{m,{\rm Newt}}$ is almost constant, while the relativistic function $h_{1,{\rm GR}}=r^{-m}e^{(m-1)\nu}T_{m,{\rm GR}}$ is not constant and varies throughout the star, as shown in Fig. \ref{h1plot}. As a result, we have $|dh_{1 ,{\rm GR}}/dr|\gg |dh_{1, {\rm Newt}}/dr|$ and $|h_{2, {\rm GR}}|\gg |h_{2, \rm Newt}|$. This observation combined with \eqref{EdotZetaRmode} and \eqref{EdotDRmode}, explains why $\tau_{\mathcal{D}, {\rm GR}}\ll \tau_{\mathcal{D}, {\rm Newt}}$ and $\tau_{\zeta, {\rm GR}}\ll \tau_{\zeta, {\rm Newt}}$.

The discovered amplification of the dissipation due to diffusion turns out to be so strong, that (for relativistic $r$-modes) in not too hot superconducting neutron stars the diffusion becomes very efficient dissipative mechanism, significantly exceeding shear viscosity, which is currently believed to be one of the main sources of the $r$-mode dissipation. This is illustrated in Fig.\ \ref{timescalesGRvsNewt}, where we plot the ratios $\tau_{\eta}/\tau_\mathcal{D}$ for the relativistic (shown in blue) and Newtonian (shown in red) $r$-modes. Since $\tau_{\eta}\propto (T^{\infty})^2$ and $\tau_{\mathcal{D}}\propto (T^{\infty})^2$ with high accuracy, the ratio $\tau_\eta/\tau_\mathcal{D}$ is almost temperature-independent.  We see that, while in the Newtonian theory shear viscosity is, indeed, stronger than the diffusion, in GR diffusion dominates in the whole range of the considered rotation rates. Therefore, we expect that diffusion will significantly affect the shape of the relativistic $r$-mode instability window at not too high stellar temperatures.


\begin{figure*}
\begin{minipage}{0.5\linewidth}
\centering\includegraphics[width=1.0\linewidth]{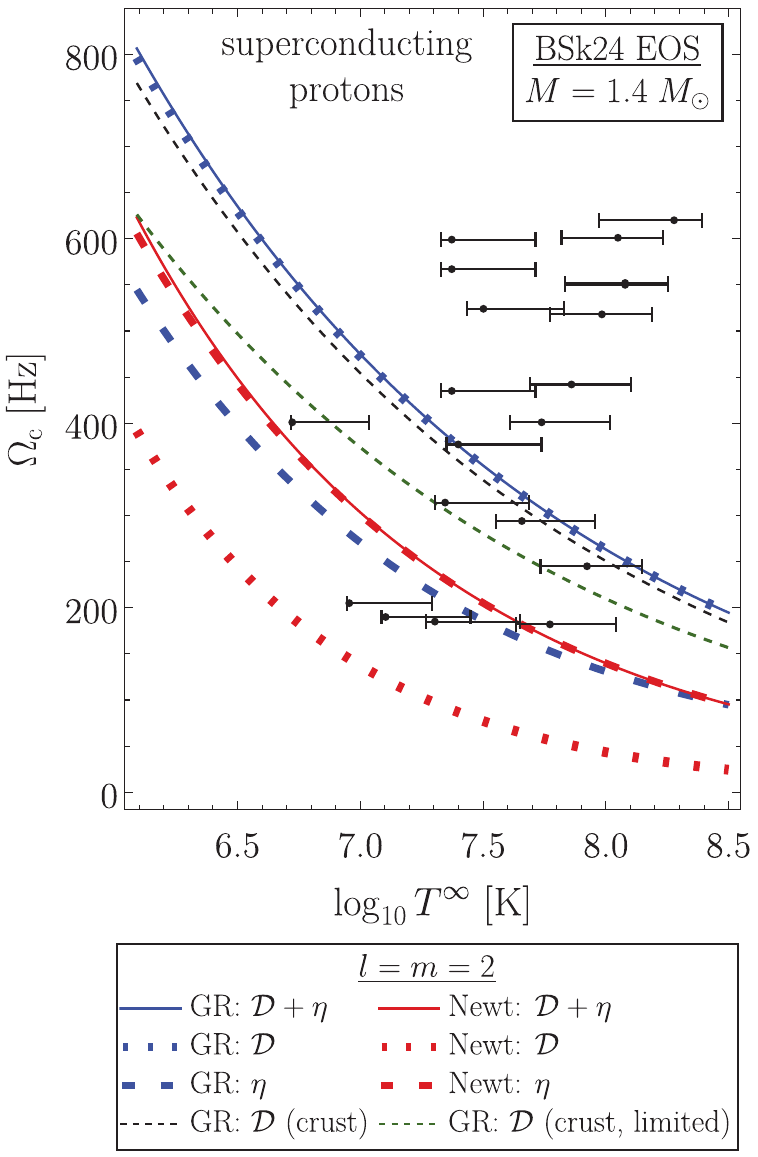}
\end{minipage}
\hfill
\begin{minipage}{0.48\linewidth}
\caption{Instability windows of the relativistic (``GR'') and Newtonian (``Newt'') fundamental $l=m=2$ $r$-modes in superconducting NS. Results, obtained for completely nonbarotropic stellar matter, are shown in red and blue by solid lines (stabilization by diffusion and shear viscosity,``$\mathcal{D}+\eta$''), short thick dashes (stabilization only by diffusion,``$\mathcal{D}$''), and long thick dashes (stabilization only by shear viscosity,``$\eta$''). Thin black dashes [``$\mathcal{D}$ (crust)''] show the relativistic $r$-mode stabilization by diffusion in an NS with nonbarotropic core and barotropic crust, while thin green dashes [``$\mathcal{D}$ (crust, limited)''] show the stabilization by diffusion in the very same NS but with proton superconductivity limited to the core region with density $\rho\leq 4\times 10^{14}{\rm g / cm^3}$ (which approximately corresponds to $0.76<r/R<0.92$).  Each curve divides the $(\Omega, T^\infty)$-plane into the upper area, where the modes are unstable, and the lower area, where they are damped by dissipative mechanisms. Black dots and bars represent the observational data on LMXBs taken from \cite{kgd2021}. The error bars represent the uncertainty in our knowledge of the internal temperature associated with the poorly known composition of the outer NS layers (see \cite{gck2014} for details). }
\label{windowplot}
\end{minipage}
\end{figure*}



\begin{figure*}
\begin{minipage}{0.48\linewidth}
\caption{Relativistic (``GR'', solid lines) and Newtonian (``Newt'', dashes) fundamental $l=m=2$ $r$-mode instability windows for the case of normal $npe\mu$-matter. Calculations with bulk viscosity, associated with modified Urca processes (``mUrca''), are shown in blue. Those with bulk viscosity due to the  direct Urca process are shown in red (``dUrca 1/2'''; the process is allowed in the region $0\leq r\leq R/2$) and green (``dUrca''; the process is allowed in the entire neutron star). Black dots and bars represent the observational data, see caption to Fig.\ \ref{windowplot} for details.
}
\label{windowplotNorm}
\end{minipage}
\hfill
\begin{minipage}{0.5\linewidth}
\includegraphics[width=1.0\linewidth]{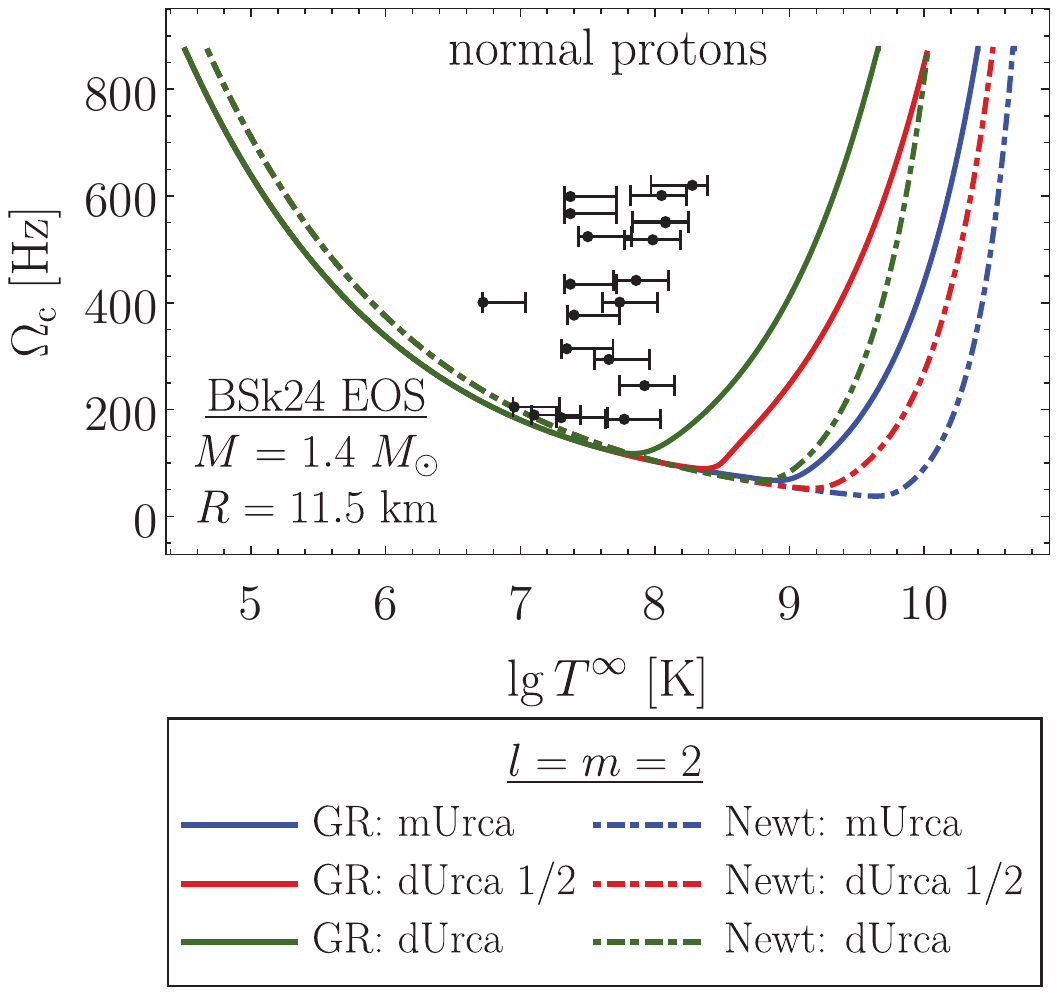}
\end{minipage}

\end{figure*}


To investigate the combined effects of the $r$-mode nonanalyticity and particle diffusion in strongly superconducting matter on the $r$-mode instability window, we consider three stabilization scenarios, each leading to a corresponding critical angular velocity, $\Omega_{\rm c}(T^\infty)$, at which the mode amplification by CFS mechanism equals the mode damping by dissipative mechanisms of interest:
\begin{enumerate}

\item[1)]
mode stabilization by shear viscosity:
\begin{gather*}
\langle\dot{\tilde{E}}_{\rm GW}(\Omega_{\rm c})+\dot{\tilde{E}}_{\eta}(\Omega_{\rm c},T^\infty)\rangle_P=0,
\end{gather*}

\item[2)]
mode stabilization by diffusion:
\begin{gather*}
\langle\dot{\tilde{E}}_{\rm GW}(\Omega_{\rm c})+\dot{\tilde{E}}_{\mathcal{D}}(\Omega_{\rm c},T^\infty)\rangle_P=0,
\end{gather*}

\item[3)]
mode stabilization by shear viscosity and diffusion:
\begin{gather*}
\langle\dot{\tilde{E}}_{\rm GW}(\Omega_{\rm c})+\dot{\tilde{E}}_{\eta}(\Omega_{\rm c},T^\infty)+\dot{\tilde{E}}_{\mathcal{D}}(\Omega_{\rm c},T^\infty)\rangle_P=0.
\end{gather*}
\end{enumerate}

We show the resulting critical curves, $\Omega_{\rm c}(T^\infty)$, calculated for relativistic (``GR'') and Newtonian (``Newt'') fundamental $l=m=2$ $r$-modes in Fig. \ref{windowplot}. Each curve divides the $(\Omega,T^\infty)$-plane into the upper area, where the modes are unstable, and the lower area, where they are damped by dissipative mechanisms of interest. Apart from the completely nonbarotropic neutron star model, introduced above (the corresponding results are shown in red and blue), we have also considered the relativistic $r$-mode stabilization by diffusion in a more realistic stellar model, accounting for the barotropic NS crust (see black and green thin dashes; the crust model is the same as in Ref.\ \cite{kgk2022_1}). Black dashes show the instability window, calculated assuming that the strong proton superconductivity resides in the whole NS core. Green dashes show the estimate of the $r$-mode instability window in the case when proton superconductivity weakens at large densities, as predicted by some microscopic theories (see, e.g., Refs.\ \cite{bs2007, guo_etal19,sc19,lh21}) and is only active in the core up to the density $\rho\lesssim 4\times 10^{14}$~g~cm$^{-3}$ (see Fig.\ 7 in Ref.\ \cite{Hoetal2015}), which approximately corresponds to the region $0.76<r/R<0.92$ in our model. As expected, we see that in the Newtonian theory the effect of particle diffusion on the $r$-mode instability window is rather weak, and the latter is determined primarily by dissipation due to shear viscosity, whereas in general relativity the picture is exactly the opposite: particle diffusion determines the shape of the window and the effect of shear viscosity is negligibly small, even if diffusion is only efficient in the outer layers of the stellar core (see the green dashed curve).

Black dots with error bars in the figure represent the observational data on NSs in LMXBs, taken from Ref.\ \cite{kgd2021}. We see that, unfortunately, diffusion is not sufficient to stabilize (i.e., ``explain'') all the observed sources and, therefore, inclusion of additional dissipative mechanisms (such as, e.g.,  dissipation in Ekman layer \cite{bu2000, lu2001, ga2006}, mutual friction \cite{hap2009}, resonant $r$-mode stabilization by superfluid modes \cite{kgd2020, kgd2021} and enhanced bulk viscosity in hyperonic matter \cite{nb2006, ofengeimetal2019}) is still required.

Finally, we would like to finish this section with the discussion of how the discovered amplification of the dissipation due to bulk viscosity might affect the $r$-mode instability window at high temperatures, $T^\infty \gtrsim 10^9 \ {\rm K}$. As discussed above, it is likely that at temperatures, where the bulk viscosity is efficient, the protons are not completely superconducting. For this reason (and having in mind application to hyperon stars), below we, for simplicity, restrict ourselves to the case of normal matter. In the absence of proton superconductivity particle diffusion is inefficient (for both Newtonian and relativistic $r$-modes) and shear viscosity (to be adopted from \cite{ss2018}) usually becomes the leading dissipative mechanism at $T^\infty\lesssim 10^9 \ {\rm K}$, while at higher temperatures the energy dissipation is mainly caused by the bulk viscosity. In addition to the bulk viscosity associated with the modified Urca processes, we will also, for illustrative purposes, consider the bulk viscosity associated with the direct Urca process (taken from \cite{haenseletal2000}). Although the latter is not allowed in our model, we can formally consider two scenarios, when this process proceeds either in the whole star or, say, in its inner half, $0\leq r\leq R/2$.  In this way, ignoring dissipation due to diffusion, we obtain the instability windows shown in Fig.\ \ref{windowplotNorm}.

We see that, independently of the processes behind the bulk viscosity (modified or direct Urca), at low temperatures the relativistic and Newtonian instability windows almost coincide, while at high temperatures relativistic instability curve goes higher than Newtonian one, reflecting the fact that in GR dissipation due to bulk viscosity is significantly amplified compared to the Newtonian case. This amplification, however, is not strong enough to stabilize the observed sources and we again arrive at a conclusion, that the theory requires additional dissipative mechanisms. In particular, relativistic (i.e., accounting for the discussed amplification) consideration of the dissipation due to strong bulk viscosity in hyperonic matter \cite{nb2006, ofengeimetal2019} is an interesting problem to be addressed in the subsequent publication.


\section{Discussion}\label{IV Conclusion}

Until recently, calculating the evolutionary timescales and instability windows of relativistic $r$-modes in nonbarotropic neutron stars was a challenging task due to their unclear properties and the lack of a viable and consistent method to determine their oscillation spectrum and eigenfrequencies (the problem of the continuous $r$-mode spectrum mentioned above). However, our discovery of the nonanalytic (in $\Omega$) behavior of relativistic $r$-modes  \cite{kgk2022_1, kgk2022_2} has allowed us to overcome these difficulties. We have obtained relativistic $r$-mode eigenfunctions (with certain simplifying assumptions) and proceeded with investigating the impact of various evolutionary mechanisms on relativistic $r$-modes, in comparison to their Newtonian counterparts.

In addition to studying the evolution of $r$-modes influenced by the commonly considered CFS instability and dissipative effects from shear and bulk viscosities, we have also explored the role of diffusion as a dissipative mechanism for relativistic $r$-modes. Initially, our motivation for including diffusion in the study was based on the significant energy losses observed in certain stellar oscillation modes (such as sound waves, $p$-modes, and $g$-modes) that exist in neutron stars with strong proton superconductivity in their cores. These losses surpass those attributed to shear viscosity, which is typically considered a primary dissipative agent in moderately hot neutron stars. Although our preliminary estimates \cite{kgk2021} suggested that this might not be the case for $r$-modes, these estimates did not fully account for the newly discovered nonanalytic behavior of relativistic $r$-modes in nonbarotropic matter.

We began this paper by discussing the general theoretical framework used to study oscillations and their evolution in nonbarotropic strongly superconducting stellar matter, considering the influence of CFS instability (GW), shear viscosity ($\eta$), bulk viscosity ($\zeta$), and diffusion ($\mathcal{D}$). Except for the diffusion
\footnote{The developed framework can not be directly applied to diffusion in normal matter because the expression for the self-consistent electric field in normal matter differs from Eq.\ (\ref{Efield}).
},
this framework can also be applied to oscillations in normal matter, with the only difference being the replacement of the coefficients $\eta$ and $\zeta$ with their normal matter counterparts.

We then focused on applying this framework to the case of relativistic $r$-modes. We derived explicit expressions \eqref{ETildeRmode}--\eqref{EdotGWRmode} for the energy $\tilde{E}$ of $r$-modes and their (averaged over the oscillation period) energy change rates $\langle\dot{\tilde{E}}_{\zeta,\eta,\mathcal{D},{\rm GW}}\rangle_P$. These expressions are valid for slowly rotating neutron stars in GR and, with minor modifications, in Newtonian theory. Our analysis of these expressions revealed
that at slow rotation rates, relativistic $r$-modes, due to their nonanalytic behavior, are less affected by the CFS instability compared to their Newtonian counterparts. However, the effects of dissipative mechanisms in GR are expected to be significantly stronger than in the Newtonian case.

To illustrate these differences, we considered the limit of extremely slow stellar rotation and examined how exactly the relativistic $r$-mode nonanalyticity influences the dependence of their evolutionary timescales $\tau_{\zeta,\eta,\mathcal{D},{\rm GW}}=(2\tilde{E})/|\langle\dot{\tilde{E}}_{\zeta,\eta,\mathcal{D},{\rm GW}}\rangle_P|$ on $\Omega$.

These findings received further confirmation in our numerical calculations. Employing a simplified model of a neutron star, consisting of a liquid inner $npe\mu$-core and outer $npe$-core, we determined the evolutionary timescales of the Newtonian and relativistic $l=m=2$ fundamental $r$-modes. The results not only confirmed our theoretical predictions but also revealed an unexpected finding: for relativistic $r$-modes, diffusion and bulk viscosity are significantly more effective damping agents than for Newtonian $r$-modes, even at high rotation rates.

Our analysis indicates that this discrepancy is not primarily caused by the relativistic $r$-mode nonanalyticity but rather by the distinct behavior of the Newtonian $r$-mode eigenfunctions. Unlike in general relativity, in Newtonian theory the function $h_{1,{\rm Newt}}$ \eqref{h1func} is almost constant and, therefore the functions, which define $\dot{\tilde{E}}_{\zeta,\mathcal{D}}$ \eqref{EdotZetaRmode}, \eqref{EdotDRmode}, satisfy $|h_{1,{\rm Newt}}'(r)|\ll |h_{1,{\rm GR}}'(r)|$, $|h_{2,\rm Newt}|\ll |h_{2,\rm GR}|$ and $|h_{2,{\rm Newt}}'(r)|\ll |h_{2,{\rm GR}}'(r)|$.
This behavior of Newtonian eigenfunctions, leads to the suppression of all the terms contributing to $\langle\dot{\tilde{E}}_{\zeta,\mathcal{D}}\rangle_P$, even at high rotation rates, in contrast to their relativistic counterparts. The resulting disparity in diffusive dissipation is so significant that in the case of superconducting matter, diffusion becomes the dominant dissipative mechanism for relativistic $r$-modes, surpassing the dissipation caused by shear viscosity by several orders of magnitude.

Finally, for the considered toy model of a neutron star, we have calculated the instability windows for both relativistic and Newtonian $r$-modes. At not too high temperatures, we have explored three scenarios for stabilizing the $r$-modes: shear viscosity alone, diffusion alone, or a combination of diffusion and shear viscosity. As predicted, the shape of the instability window for relativistic $r$-modes is primarily governed by diffusion, with shear viscosity playing a negligible role. This stands in stark contrast to the behavior observed for Newtonian $r$-modes. Furthermore, to evaluate the impact of bulk viscosity amplification in general relativity on the instability windows, we intentionally disabled proton superconductivity and computed the instability window for both relativistic and Newtonian $r$-modes stabilized by shear and bulk viscosities (disregarding diffusion in normal matter). As anticipated, at high temperatures where the window's shape is primarily determined by bulk viscosity, relativistic $r$-modes exhibit significantly more efficient stabilization compared to their Newtonian counterparts.

Summarizing, the main result of this study is that accounting for GR and particle diffusion in nonbarotropic stellar matter may significantly affect the $r$-mode instability windows. At low and moderate stellar temperatures in the presence of proton superconductivity diffusion becomes the leading dissipative mechanism, much stronger than shear viscosity. Considering normal (nonsuperconducting) matter we have also shown that at high temperatures relativistic $r$-modes lose energy due to bulk viscosity much faster than their Newtonian cousins. A number of comments, concerning the validity of the made approximations, however, should be made.

First, in our consideration we (almost) completely ignore the effects, associated with the neutron star crust. To justify this approximation, we note that diffusion and bulk viscosity are not very efficient in the crust (but see \cite{ygh18}). Being rather thin compared to the neutron star core \cite{kgk2022_1}, the crust will only slightly affect the $r$-mode eigenfunctions at high rotation rates, typical for the instability windows. Indeed, the $r$-mode eigenfunctions  are  practically not suppressed in the core at these rotation rates so that, for instance, the dissipation due to shear viscosity in a thin crust cannot lead to significant additional losses of mechanical energy. Moreover, according to our calculations (see Fig. \ref{windowplot}), critical curves, corresponding to the relativistic $r$-mode stabilization by diffusion in NS models with and without crust are close to each other, and thus accounting for the crust does not qualitatively change the $r$-mode instability window. At small $\Omega$, however, instead of being localized in the vicinity of the stellar surface, eigenfunctions will localize in the vicinity of the crust-core interface and inside the crust. Therefore, at extremely slow rotation rates the crust will provide the leading contribution to the $r$-mode energy \eqref{ETildeRmode} and the $r$-mode energy change rate \eqref{EdotGWRmode} due to CFS-instability, since the corresponding subintegral expressions do not contain large eigenfunction gradients, while eigenfunctions themselves are suppressed almost everywhere in the core. As a result, the dependence of the $r$-mode evolutionary timescales on $\Omega$ differ from those in the purely nonbarotropic neutron star. Particularly, $\tau_{\rm GW}$ recovers its traditional dependence on the rotation rate, $\tau_{\rm GW}\propto \Omega^{2m+2}$. Nevertheless, in our numerical calculations, the CFS instability of relativistic $r$-modes is still suppressed compared to the Newtonian case for $\Omega\lesssim 0.05 \ \Omega_{\rm K}$.

Second, in our calculations we ignore the quadratic rotational corrections to the background, describing the oblateness of the neutron star due to rotation. Strictly speaking, these corrections should be accounted for in the most accurate calculation but, at the same time, Newtonian calculations show that they may affect mostly the $r$-mode eigenfrequency correction $\sigma^{(1)}$, while the $r$-mode toroidal function is practically insensitive to their inclusion. As a result, accounting for the rotational corrections to the background should not have significant effect on the Newtonian $r$-mode properties. This conclusion, in fact, is corroborated by the straightforward comparison of the obtained here Newtonian $r$-mode instability windows with the calculation accounting for the background rotational corrections, presented in our previous study \cite{kgk2021}. As for relativistic $r$-modes, accounting for these corrections will not change their most important properties -- nonanalyticity and peculiar ordering, -- responsible for their drastically different behavior under the influence of the evolutionary mechanisms, discussed in the paper. Moreover, our preliminary calculations show that accounting for the rotational corrections does not affect the expressions \eqref{EdotZetaRmode}-\eqref{EdotGWRmode} for the relativistic $r$-mode energy change rates, and, at the same time, only weakly influences relativistic $r$-mode eigenfunctions.

Third, in our calculations we employ the Cowling approximation, i.e., in the Euler equations, continuity equations and in the derivation of the energy change rate due to CFS mechanism we ignore perturbations of the gravitational field, assuming that they are small, compared to hydrodynamic ones. This approximation significantly simplifies the problem and, at the same time, is known to provide reasonable estimates of the $r$-mode properties \cite{jc2017}. Moreover, as our preliminary results indicate \cite{kgk2022_1}, the most general equations, that govern the dynamics of nonanalytic relativistic $r$-modes in the $\Omega\to 0$ limit, {\it exactly} coincide with those, obtained within the Cowling approximation in the $\Omega\to 0$ limit. Of course, in the most accurate calculation one should account for all the metric tensor perturbations and look for the complex eigenfrequency correction, that would determine the mode driving timescale, associated with the CFS instability, but such calculation deserves separate consideration and goes far beyond the scope of this work.

Fourth, in our study we completely ignore the neutron superfluidity, present in the stellar matter. As discussed in one of our previous works \cite{kgk2021}, within the conditions of neutron star interiors, we found that the weaker the friction forces acting between different particle species, the more pronounced the diffusive dissipation becomes. This pattern can be understood as follows: Strong friction forces effectively bind particle species together, suppressing their relative motion, which is the primary cause of dissipation. The efficiency of friction forces is directly influenced by the frequency of particle collisions in the stellar matter. In normal $npe\mu$-matter strong neutron-proton interaction and electromagnetic baryon-lepton and lepton-lepton interactions do not allow different particle species to acquire significant relative velocities and, therefore, dissipation due to diffusion is weak. Proton superconductivity weakens the interactions of particles with protons and, therefore, allows for larger relative neutron-lepton velocities, leading to stronger diffusive dissipation. Now, if protons are strongly superconducting and neutrons are strongly superfluid in a given stellar region (i.e., $T\ll T_{{\rm c}n}$, where $T_{{\rm c} n}$ is the neutron transition temperature to superfluid state), their scattering on other particle species is suppressed. Consequently, diffusive dissipation arises mainly due to the interaction of charged particles, which is significantly stronger than neutron-lepton interaction. As a result, diffusive dissipation in such region is suppressed (the stronger the friction forces the lower the dissipation). In other words, a stellar region can significantly contribute to dissipation due to diffusion only if it contains strongly superconducting protons and normal or weakly-superfluid neutrons. Here we would like to note that even when diffusive dissipation is confined to such regions, it still may (depending on the superfluidity model) be strong enough to surpass that due to shear viscosity. Although the theoretical framework for the calculation of diffusion coefficients in mixtures of superfluid/superconducting Fermi-liquids has recently been developed \cite{gg2023}, the detailed calculations have not yet been performed, so at this point we can only make estimates by varying the stellar regions contributing to diffusive dissipation. For instance, as illustrated in Fig.\ \ref{windowplot}, even when proton superconductivity is confined to a narrow region near the crust-core interface, the relativistic $r$-mode instability window is still primarily determined by diffusion rather than shear viscosity. In addition to its impact on microphysics, neutron superfluidity also influences the hydrodynamic equations and introduces mutual friction as an additional channel for the leakage of $r$-mode energy. Detailed study of these effects on the $r$-mode instability window requires separate consideration.

Lastly, several Newtonian studies \cite{yl2000, ag2023, ga2023} suggest that, at high rotation rates, $r$-modes may effectively violate the assumed ordering and behave as if the matter were barotropic, due to weak matter nonbarotropicity (i.e., a small Schwarzschild discriminant). However, in Appendix \ref{AppD}, we address this issue in the context of relativistic $r$-modes and demonstrate that, at least for the fundamental $l=m=2$ $r$-mode considered in this study, this is not the case.

In conclusion, we must acknowledge that while diffusion and bulk viscosity are indeed significantly more efficient for stabilizing relativistic $r$-modes compared to Newtonian ones, they are still not strong enough to stabilize all observed sources. This suggests that the developed theory requires the inclusion of additional dissipative mechanisms. Some potential candidates include, but not limited to, dissipation in the Ekman layer \cite{bu2000, lu2001, ga2006}, resonant $r$-mode stabilization by superfluid modes \cite{kgd2020, kgd2021}, enhanced bulk viscosity in hyperonic matter \cite{nb2006, ofengeimetal2019}, and vortex-mediated mutual friction \cite{hap2009}. To obtain a comprehensive understanding of the properties of relativistic $r$-modes, all these mechanisms should be taken into account in the general calculation of the instability window. 
Addressing this intriguing problem can provide valuable insights into the role of $r$-modes in the dynamics of neutron stars.

\section*{Acknowledgements}
This work was supported by Russian Science Foundation [Grant 22-12-00048].
MEG and EMK are grateful to the Department of Particle Physics \& Astrophysics at the Weizmann Institute of Science for hospitality and excellent working conditions.

\appendix


\section{Oscillation energy} \label{AppA}

To find the oscillation energy $\tilde{E}$ in the corotating reference frame, we start with the perturbed form of the conservation law, associated with the Killing vector $k^\rho=u^\rho/\Lambda$:
\begin{gather}
\label{Elaw}
\nabla_\mu (k_\rho \hat{\delta}T^{\mu\rho})=0,
\end{gather}
where $\hat{\delta}f$ is the {\it exact} deviation of the quantity $f$ from the equilibrium (do not confuse with Eulerian perturbation $\delta f$). The exact four-velocity normalization condition and continuity equations can be written as
\begin{gather}
2 u_\rho \hat{\delta} \mathfrak{u}^\rho+\hat{\delta} \mathfrak{u}_\rho \hat{\delta} \mathfrak{u}^\rho=0, \\
\nabla_\mu(\hat{\delta}n_k u^\mu+n_{k0}\hat{\delta} \mathfrak{u}^\mu+\hat{\delta}n_k\hat{\delta} \mathfrak{u}^\mu)=0.
\end{gather}
Note that, using these equations, equilibrium conditions \eqref{equilibriumMU2} and quasineutrality condition, $e_k n_k=0$, it is easy to show that the vector
\begin{gather}
F^\mu\equiv\frac{\mu_{k0}}{\Lambda}\hat{\delta} j_k^\mu=\frac{\mu_{k0}}{\Lambda}\biggl[\hat{\delta}n_k u^\mu+n_{k0}\hat{\delta} \mathfrak{u}^\mu+\hat{\delta}n_k\hat{\delta} \mathfrak{u}^\mu\biggr]
\end{gather}
is divergenceless, $\nabla_\mu F^\mu=0$. This result implies that the integral
\begin{gather}
\label{zeroF}
\mathcal{F}\equiv\int F^t\sqrt{-{\rm g}}  d^3x=\const
\end{gather}
is conserved. Moreover, using the perturbed quasineutrality condition, $e_k \hat{\delta}j_k^t=0$, and chemical equilibrium condition \eqref{equilibriumMU2} with $m={\rm s}$ it is easy to see, that it vanishes:
\begin{multline}
\mathcal{F}=\int \frac{\mu_{k0}}{\Lambda}\hat{\delta}j_k^t \sqrt{-{\rm g}} d^3x= \\
=\int \frac{\mu_{k0}-e_k/e_{\rm s} \mu_{{\rm s}0}}{\Lambda}\hat{\delta}j_k^t \sqrt{-{\rm g}} d^3x= \\
=\frac{\mu_{k0}-e_k/e_{\rm s} \mu_{{\rm s}0}}{\Lambda}\int \hat{\delta}j_k^t \sqrt{-{\rm g}} d^3x=\\
=\frac{\mu_{k0}}{\Lambda}\hat{\delta}N_k=0.
\end{multline}
Here in the final step, we have utilized the conservation of the total number of particle species, ensuring that its variation vanishes in the perturbed star,
\begin{gather}
\hat{\delta}N_k=\int \hat{\delta}j_k^t \sqrt{-{\rm g}} d^3x=0.
\end{gather}

Now, retaining the terms up to the second order in $\hat\delta$, we obtain
\begin{multline}
k_\rho\hat{\delta} T^{\mu\rho}=-\frac{u^\mu}{\Lambda}\biggl[\frac{w_0}{2}\hat{\delta} \mathfrak{u}_\rho \hat{\delta} \mathfrak{u}^\rho+\hat{\delta}\mu_k n_{k0}+\hat{\delta}\mu_k \hat{\delta}n_k-\hat{\delta}p\biggr]- \\
-\frac{1}{\Lambda}n_{k0}\hat{\delta}\mu_k \hat{\delta} \mathfrak{u}^\mu-F^\mu.
\end{multline}
Within the required accuracy, everywhere in the second $\hat\delta$-order terms one may safely replace $\hat{\delta}f$ with the solutions of the linearized equations $\delta f$. For the first order terms, however, it is necessary to use the following relations, accurate up to the second order:
\begin{gather}
\begin{gathered}
\hat{\delta}p\approx n_{n0}\biggl(\frac{\p \mu_n}{\p n_k}\biggr)_0\hat{\delta}n_k+\frac{1}{2}\biggl(\frac{\p}{\p n_m}n_n\frac{\p \mu_n}{\p n_k}\biggr)_0\delta n_k \delta n_m, \\
\hat{\delta}\mu_n\approx \biggl(\frac{\p\mu_n}{\p n_k}\biggr)_0\hat{\delta}n_k+\frac{1}{2}\biggl(\frac{\p^2 \mu_n}{\p n_k \p n_m}\biggr)_0\delta n_k \delta n_m. \raisetag{1cm}
\end{gathered}
\end{gather}
Using these formulas and linearized thermodynamic relations, we find:
\begin{multline}
k_\rho\hat{\delta}T^{\mu\rho}\approx-\frac{u^\mu}{2\Lambda}\biggl[w_0\delta\mathfrak{u}_\rho\delta\mathfrak{u}^\rho+ \delta\mu_k \delta n_k \biggr]-\\
-\frac{1}{\Lambda}\delta p\delta\mathfrak{u}^\mu-F^\mu.
\end{multline}
Now, integrating the conservation law \eqref{Elaw} using the fact that the integral \eqref{zeroF} over $F^t$ vanishes and that $u^t=\Lambda$, we find the expression for the energy in the corotating reference frame:
\begin{gather}
\tilde{E}=\int\biggl[\frac{w_0}{2}\delta\mathfrak{u}_\rho \delta\mathfrak{u}^\rho+\frac{1}{2}\delta\mu_k\delta n_k+\frac{1}{\Lambda}\delta p\delta\mathfrak{u}^t\biggr]\sqrt{-{\rm g}} \ d^3 x.
\end{gather}
Finally, using thermodynamic relations and general relations between the Eulerian perturbations, one can show that
\begin{gather}
\delta\mu_k \delta n_k=\frac{(\delta p)^2}{\gamma p_0}+w_0(\xi^\rho \mathcal{A}_\rho)(\xi^\lambda\nabla_\lambda \ln\Lambda),
\end{gather}
so that the final expression for the energy takes the form
\begin{multline}
\tilde{E}=\int\biggl[\frac{w_0}{2}\delta\mathfrak{u}_\rho \delta\mathfrak{u}^\rho+\frac{1}{2}\frac{(\delta p)^2}{\gamma p_0}+\frac{1}{\Lambda}\delta p \delta\mathfrak{u}^t+\\
+\frac{1}{2}w_0(\xi^\rho \mathcal{A}_\rho)(\xi^\lambda\nabla_\lambda \ln\Lambda)\biggr]\sqrt{-{\rm g}} \ d^3 x.
\end{multline}
%


\section{Gravitational multipole moments}\label{AppB}

As mentioned in the main text, to compute the energy loss rate due to the emission of gravitational waves, we use the general formalism, developed by Thorne \cite{thorne1980}. His mathematical notation, as he admits, is ``somewhat special", and it is our feeling that it should be briefly reviewed. In his analysis, Thorne uses the ``physical" basis $\hat{{\bf e}}_{\mu}$ instead of the conventional coordinate basis ${\bf e}_\mu\equiv\p_\mu$:
\begin{gather}
\hat{{\bf e}}_{\mu}=S_{\mu}{}^\rho {\bf e}_\rho, \quad {\bf e}_{\mu}=(S^{-1}){}_{\mu}{}^{\rho} \hat{{\bf e}}_{\rho}, \\
S_{\mu}{}^\rho=\diag\{1,1,1/r,1/r\sin\theta\}.
\end{gather}
Thorne also treats coordinates as though the spacetime is flat and uses the ``flat" metric tensor $\eta_{\mu\nu}\equiv\diag\{-1,1,1,1\}$ to raise and lower tensor indices. We have to note that such notation leads to ambiguities in definitions of tensor components. For example, let us consider the vector field with components $V^\mu$ with respect to the basis ${\bf e}_\mu$. On one hand, its components $\hat{V}^\mu$ with respect to the basis $\hat{\bf e}_\mu$ can be found as $\hat{V}^{\rho}=V^\mu (S^{-1})_\mu{}^{\rho}$, and we then obtain $\hat{V}_{\mu}=V^{\rho}(S^{-1}){}_{\rho}{}^{\kappa}\eta_{\kappa\mu}$. On the other hand, we can first lower the index with ${\rm g}_{\mu\nu}$, then perform the transition to the physical basis, and find $\hat{V}_{\mu}=S_{\mu}{}^\rho {\rm g}_{\rho\kappa}V^{\kappa}$, which, generally, differs from the previous result. In order to avoid such ambiguities it is necessary to state, which components, $V^\mu$ or $V_\mu$, should be viewed as primary.

Following Thorne [see Eq.\ (5.3) in Ref.\ \cite{thorne1980}], we consider the contravariant components $\tau^{\mu\nu}$ of the effective stress-energy pseudotensor  of the whole system ``neutron star + gravitational radiation" as primary, and then define $\hat{\tau}^{\mu\nu}=\tau^{\eta\kappa}(S^{-1}){}_{\eta}{}^{\mu}(S^{-1}){}_{\kappa}{}^{\nu}$ and $\hat{\tau}_{\mu\nu}=\eta_{\mu\rho}\eta_{\nu\lambda}\hat{\tau}^{\rho\lambda}$. Particularly, within the Cowling approximation, we will use  $\hat{\tau}^{\mu\nu}\approx(-\hat{{\rm g}})\hat{T}^{\mu\nu}$, where ${\hat{\rm g}}=\det \hat{{\rm g}}_{\mu\nu}$. Next, we introduce the spherical harmonic ${Y}^{LM}(\theta,\varphi)$ and the magnetic-type vector-spherical harmonic ${\bf Y}^{{\rm B}, LM}(\theta,\varphi)$ defined as (in the physical basis):
\begin{gather}
{\bf Y}^{{\rm B}, LM}=\frac{1}{\sqrt{L(L+1)}}\biggl\{0,-\frac{1}{\sin\theta}\frac{\p{Y}_{LM}}{\p\varphi}, \frac{\p{Y}_{LM}}{\p\theta} \biggr\}.
\end{gather}
In these notations, mass $\mathcal{I}_{LM}$ and current $\mathcal{S}_{LM}$ multipole moments, required in the calculation of $\dot{E}_{\rm GW}$ \eqref{EdotGW}, are given by the following formulas
\begin{gather}
\begin{gathered}
\mathcal{I}_{LM}=\int\hat{\tau}_{tt}{Y}^{LM (\star)} r^L r^2 \sin\theta \ d^3x, \\
\mathcal{S}_{LM}=2\sqrt{\frac{L}{L+1}}\int(-\hat{\tau}_{tk})Y^{{\rm B},LM (\star)}_{k} r^L r^2 \sin\theta \ d^3x, \raisetag{1.25cm}
\end{gathered}
\end{gather}
where $d^3x=dr \ d\theta \ d\varphi$, ``$\star$" is the complex conjugation, and the summation over the spatial $k$ index is implied.


\section{Coefficients in the r-mode equations} \label{AppC}

Let us introduce the following notations:
\begin{gather}
\begin{gathered}
k^+_L=\frac{L+m}{2L+1}, \quad
k^-_L=\frac{L-m+1}{2L+1}, \\
\gamma_1=\frac{m^2}{(m+1)^2(2m+3)}, \quad
\gamma_2=\frac{m^3[9-m(3m+2)]}{(m+1)(2m+3)}, \\
F=2 r \biggl[\lambda'-g\biggl(\frac{ c }{c_s}\biggr)^2\biggr]+5, \quad
g=-\frac{p_0'}{w_0}=\nu'.\raisetag{1cm}
\end{gathered}
\end{gather}
Then the coefficients, that appear in the system \eqref{r-modes sys} and equations \eqref{eigenfunctions}, can be written as:
\begin{widetext}
\begin{gather}
\begin{gathered}
C_1(r)=-\frac{2rk^+_{m+1}}{(m+1)^2}, \quad C_2(r)=\frac{k^{+}_{m+1}}{(m+1)^2}\biggl[2rg(m+1)-F-2m-1\biggr], \quad C_3(r)=\frac{r^2 e^{-2\nu}}{c^2 m^2(m+1)^2}\biggl[\gamma_2+8 m\gamma_1 \frac{c^2}{c_s^2}\biggr] \\
G_1(r)=A+g(m-1)-\frac{m}{r}, \quad G_2(r)=\frac{Ac^2g(m+1)^2 e^{2\nu}}{4 \ r \ m \ k^-_m}, \quad
\Pi(r)=-\frac{4 r m e^{-2 \nu} w_0}{c^2 (m+1)^2 (2 m+1)}, \\
q_1=\frac{(2 m+3)(m+1)}{2 (m+2) (2 m+1)}, \quad
q_3(r)=-\frac{4 r m (r g-1) e^{-2 \nu}}{c^2 (m+1)^2 (m+2)(2m+1) A g}, \\
q_2(r)=-\frac{m e^{-2 \nu} \left\{r A \left[r (m-1)^2 g-8\right]+8 (r g-1) [r (m-1) g-m]\right\}}{2 c^2 (m+1)^2 (m+2) (2 m+1) A g}, \\
t_1=\frac{(m+1)^3 (m+3) }{2 (m+2) (2 m+1) (2 m+3)}, \quad
t_3(r)=-\frac{4 r m (m+3) [r g-1] e^{-2 \nu}}{c^2 (m+1) (m+2) (2 m+1) (2 m+3) A g}, \\
t_2(r)=-\frac{m e^{-2 \nu} \left\{r A \left[rg\left(m \left[m (m+5)+7\right]+11\right)-8 (m+3)\right]+8 (m+3)(r g-1) [r (m-1) g-m]\right\}}{2 c^2 (m+1) (m+2) (2 m+1) (2 m+3) A g}. \raisetag{4cm}
\end{gathered}
\end{gather}
\end{widetext}
%


\section{Nonbarotropicity and the r-mode ordering} \label{AppD}

\begin{figure}[H]
\vspace{0.5cm}
\centering
\includegraphics[width=1.0\linewidth]{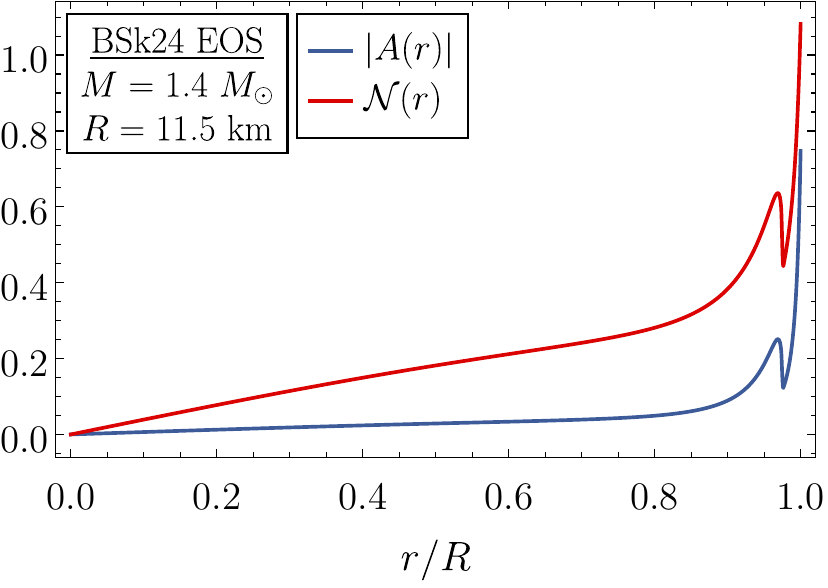}
\vspace{-0.5cm}
\caption{Dimensionless Brunt-V\"ais\"al\"a frequency (red curve) and dimensionless absolute value of the Schwarzschild discriminant (blue curve). Brunt-V\"ais\"al\"a frequency is measured in units of Keplerian velocity, $\Omega_{\rm K}\approx 1752 \ \rm Hz$.}
\label{ABVplot}
\end{figure}

\begin{widetext}

\begin{figure}[H]
\centering
\includegraphics[width=1.0\linewidth]{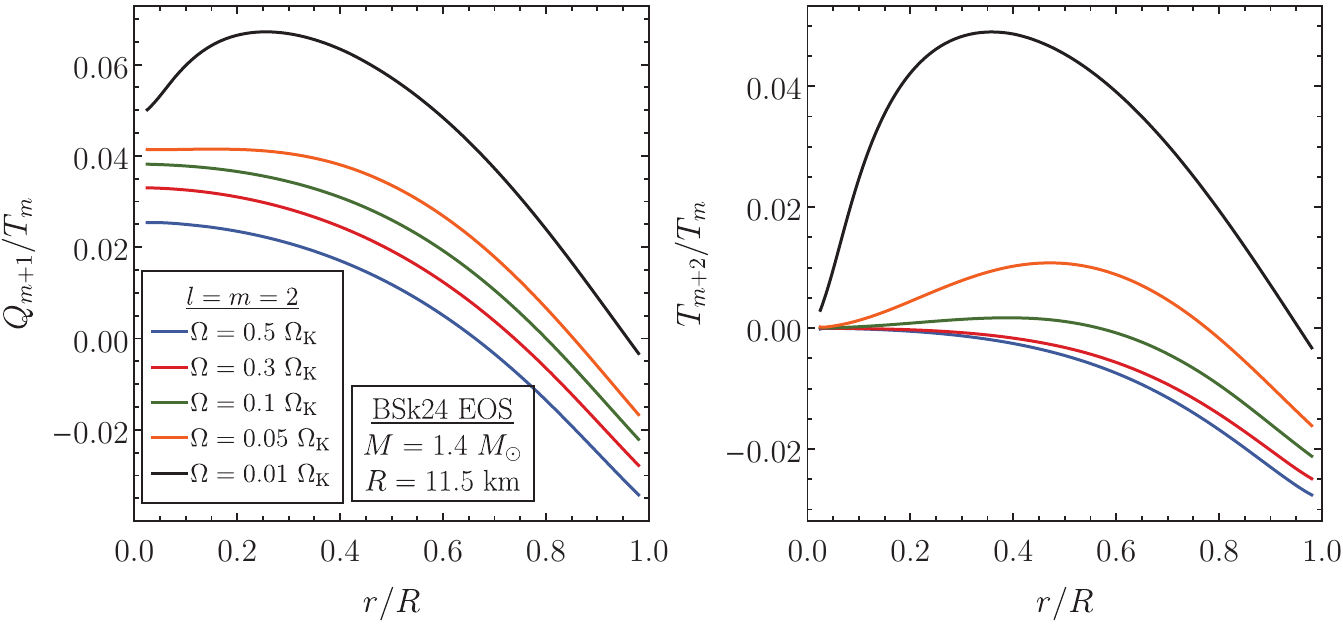}
\vspace{-0.8cm}
\caption{Ratios $Q_{m+1}/T_m$ (left panel) and $T_{m+2}/T_m$ (right panel) of the eigenfunctions of the fundamental relativistic $l=m=2$ $r$-mode, calculated for the completely nonbarotropic NS model. Different colors correspond to different stellar rotation rates.}
\label{QTplot}
\end{figure}

\end{widetext}

According to several studies \cite{yl2000, ag2023, ga2023}, at high rotation rates (i.e. those, exceeding $g$-mode eigenfrequencies), $r$-modes can effectively violate the assumed ordering and behave as if the matter were barotropic. While the analysis presented in those studies is focused solely on Newtonian $r$-modes, the argument, in principle, could be applicable to relativistic $r$-modes as well. Indeed, from appendix \ref{AppC} we see that the denominators of the coefficients $q_{2,3}$ and $t_{2,3}$ [which define the functions $Q_{m+1}$ and $T_{m+2}$, see Eq.\ \eqref{eigenfunctions}] contain the Schwarzschild discriminant $A(r)$, which is small due to weak matter nonbarotropicity. For the NS model, empoyed in this study, it is shown in Fig. \ref{ABVplot} along with the related Brunt-V\"ais\"al\"a frequency $\mathcal{N}=\sqrt{c^2 g |A|}$. Keeping in mind that $q_{2,3}$ and $t_{2,3}$ are multiplied by $\Omega^2 T_m$ and $\Omega^2 T_m'$ in \eqref{eigenfunctions}, it is reasonable to question whether the functions $Q_{m+1}$ and $T_{m+2}$ can be considered small (as we do) at rotation rates typical for the $r$-mode instability windows. Figure \ref{QTplot} addresses this question by presenting the ratios $Q_{m+1}/T_m$ and $T_{m+2}/T_m$ of the relativistic $l=m=2$ $r$-mode eigenfunctions, calculated for the completely nonbarotropic stellar model. Different colors correspond to different rotation rates, ranging from $0.5 \ \Omega_{\rm K}\approx 876 \ {\rm Hz}$  to $0.01 \ \Omega_{\rm K}\approx 18 \ {\rm Hz}$. Note that $0.5 \ \Omega_{\rm K}$ is significantly larger than $g$-mode eigenfrequencies and we, therefore, could already expect the violation of the ordering and transition to the effectively barotropic behaviour. In reality, however, we see that for all considered rotation rates the functions $Q_{m+1}$ and $T_{m+2}$ are small compared to the toroidal eigenfunction $T_m$, indicating that, at least in our model, the ordering of the considered fundamental $l=m=2$ $r$-mode is not affected by rapid rotation, and can be safely considered quasitoroidal at any reasonable rotation rate. This result resembles that of Yoshida \& Lee \cite{yl2000} for the Newtonian $l=m$ $r$-modes.


%

\end{document}